\documentclass[a4paper,fleqn,usenatbib]{mnras}

\usepackage{newtxtext,newtxmath}
\usepackage[T1]{fontenc}
\usepackage{ae,aecompl}

\usepackage{indentfirst}
\usepackage{bm}
\usepackage{hyperref}
\usepackage{float,graphicx,caption}
\usepackage[caption = false]{subfig}

\usepackage{graphicx}	% Including figure files
\usepackage{amsmath}	% Advanced maths commands
\usepackage{amssymb}	% Extra maths symbols
\defcitealias{obreschkow_2013}{O13}

\title[The triangle correlation function of phases]{Studying the morphology of reionisation with the triangle correlation function of phases}

\author[A. Gorce \& J. R. Pritchard]{
Ad\'elie Gorce,$^{1,2}$\thanks{E-mail: aeg15@ic.ac.uk (AG)}
and
Jonathan R. Pritchard,$^{1}$\thanks{E-mail: j.pritchard@imperial.ac.uk (JRP)}
\\
$^{1}$Department of Physics, Blackett Laboratory, Imperial College London, SW7 2AZ, U.K.\\
$^{2}$Institut d'Astrophysique Spatiale, CNRS/Universit\'e Paris-Sud, Universit\'e Paris-Saclay, b\^atiment 121, Universit\'e Paris-Sud, 91405 Orsay Cedex, France 
}

\date{Accepted XXX. Received YYY; in original form ZZZ}

\pubyear{2019}

\begin{document}
\label{firstpage}
\pagerange{\pageref{firstpage}--\pageref{lastpage}}
\maketitle

\begin{abstract}
We present a new statistical tool, called the triangle correlation function (TCF), inspired by the earlier work of Obreschkow et al. (2013). It is derived from the 3-point correlation function and aims to probe the characteristic scale of ionised regions during the Epoch of Reionisation from 21cm interferometric observations. Unlike most works, which focus on power spectrum, i.e. amplitude information, our statistic is based on the information we can extract from the phases of the Fourier transform of the ionisation field. In this perspective, it may benefit from the well-known interferometric concept of closure phases.
We find that this statistical estimator performs very well on simple ionisation fields. For example, with well-defined fully ionised disks, there is a peaking scale, which we can relate to the radius of the ionised bubbles. We also explore the robustness of the TCF when observational effects such as angular resolution and noise are considered. We also get interesting results on fields generated by more elaborate simulations such as 21CMFAST. Although the variety of sources and ionised morphologies in the early stages of the process make its interpretation more challenging, the nature of the signal can tell us about the stage of reionisation. Finally, and in contrast to other bubble size distribution algorithms, we show that the TCF can resolve two different characteristic scales in a given map.
\end{abstract}


%%%%%%%%%%%%%%%%%%%%%%%%%%%%%%%%%%%%%%%%%%%%%%%%%%

%%%%%%%%%%%%%%%%% BODY OF PAPER %%%%%%%%%%%%%%%%%%

\section{Introduction}

During the Epoch of Reionisation (EoR), from $z\sim 20$ to $z\sim 6$, early light sources ionised the hydrogen and helium atoms of the Intergalactic Medium (IGM). 
The information currently available on this period comes from indirect observations such as the redshift evolution of the density of star-forming galaxies, thought to be a major source of reionisation and estimated through UV luminosity densities \citep{bouwens_2015,ishigaki_2015,robertson_2015}; the integrated Thomson optical depth from Cosmic Microwave Background (CMB) observations \citep{planck_2016_reio,planck_2018_cosmo_params}; and from estimates of the averaged neutral fraction of the IGM obtained via the damping wings of quasars \citep{greig_2017,greig_2018}, surveys of Lyman-$\alpha$ emitters \citep{konno_2014,schenker_2014,mason_2018} or gamma-ray bursts afterglows \citep{totani_2014}. Although these observations keep improving, in terms of both redshift and precision, they are still not sufficient to draw a precise history of reionisation. Many uncertainties remain on various aspects of the reionisation process such as the emissivity of early galaxies or the level of clumpiness in the IGM \citep{bouwens_2017_magnitude_uncertainties,gorce_2018}. Interferometric measurements of the 21cm signal will potentially allow for maps of $\ion{H}{I}$ regions in the sky to give us a sense of both the topology and morphology of the reionisation process. To optimise the signal-to-noise ratio of current radio interferometers such as the Murchison Widefield Array (MWA)\footnote{\url{http://www.mwatelescope.org}}, the Precision Array to Probe the Epoch of Reionization (PAPER)\footnote{\url{http://eor.berkeley.edu}} and the Low Frequency Array (LOFAR)\footnote{\url{http://www.lofar.org}}, many works focus on statistical estimators such as the $n$-point correlation functions ($n$-PCF) and in particular the power spectrum ($n=2$). To make the most out of upcoming observations, especially of 21cm tomographic imaging -- a key science goal of the future Square Kilometer Array (SKA)\footnote{\url{ http://www.skatelescope.org/}}, it is useful to look at higher-order statistics. Previous works have focused on the use of the bispectrum, i.e. the Fourier transform of the 3-PCF to learn about the non-Gaussianity of the reionisation signal \citep{bispectrum_catherine_2017,bispectrum_suman_2018,bispectrum_xray_2019,trott_watkinson_2019}. Because they include both amplitude and phase information, the $n$-PCF ($n \geq 3$) will not complement the information carried by the 2-PCF as well as a correlation function solely based on the Fourier phases, such as the one we describe in this paper.

\subsection{Phase information}
\label{subsec:intro_phase_info}
 
Consider an ionisation field $x(\bm{r})$ with values ranging from 0 (neutral) to 1 (fully ionised). The visibilities observed by radio interferometers give information in Fourier space, so it is useful to expand $x(\bm{r})$ with a Fourier series,
\begin{equation}
x (\bm{r}) = \sum \ \hat{x} (\bm{k}) \ e\, ^{i\, \bm{k} \cdot \bm{r}}.
\end{equation}
Each $\hat{x}(\bm{k})$ is a complex number which can be decomposed into an amplitude $\vert \hat{x}(\bm{k}) \vert$ and a phase term $\Phi_{\bm{k}}$, so that $\hat{x}(\bm{k}) = \vert \hat{x}(\bm{k}) \vert \, \mathrm{e}^{i\, \Phi_{\bm{k}}}$. When the power spectrum $\mathcal{P}(k) = \vert \hat{x} (\bm{k}) \vert ^2$ of the field is considered, all the information contained in the phases is lost. If $x(\bm{r})$ is a pure Gaussian random field (GRF), then the phases are uniformly distributed on the interval $[0, 2\pi]$ \citep{watts_coles_2003} and the power spectrum is sufficient to fully describe the field. Conversely, non-Gaussianity is traceable to the phases. In this work, we estimate the amount of information that can be extracted from phases to add to the usual power spectrum studies. Eventually, we want to use phases to learn about the structure of an ionisation field and extract a characteristic length, such as the average radius of ionised bubbles. In this perspective, we develop a new statistical estimator, derived from the 3-point correlation function and solely based on phases, called the triangle correlation function (TCF). It is similar to the line correlation function introduced by \citet{obreschkow_2013} to study elongated structures in dark matter fields. To consider phase information only, we compute statistics from the phase factor $\hat{\epsilon} (\bm{k})$ defined as
\begin{equation}
\label{eq:def_phase_factor}
\hat{\epsilon} (\bm{k}) = \frac{\hat{x} (\bm{k})}{\vert \hat{x} (\bm{k}) \vert} = \mathrm{e}^{i \, \Phi_{\bm{k}} }.
\end{equation}
By construction, the phase factor has an amplitude of one, so that its 2-PCF and power spectrum will vanish. The simplest correlation function related to $\hat{\epsilon} (\bm{k})$ is then its 3-PCF. 

\begin{figure}
\centering
\includegraphics[width=0.4\columnwidth]{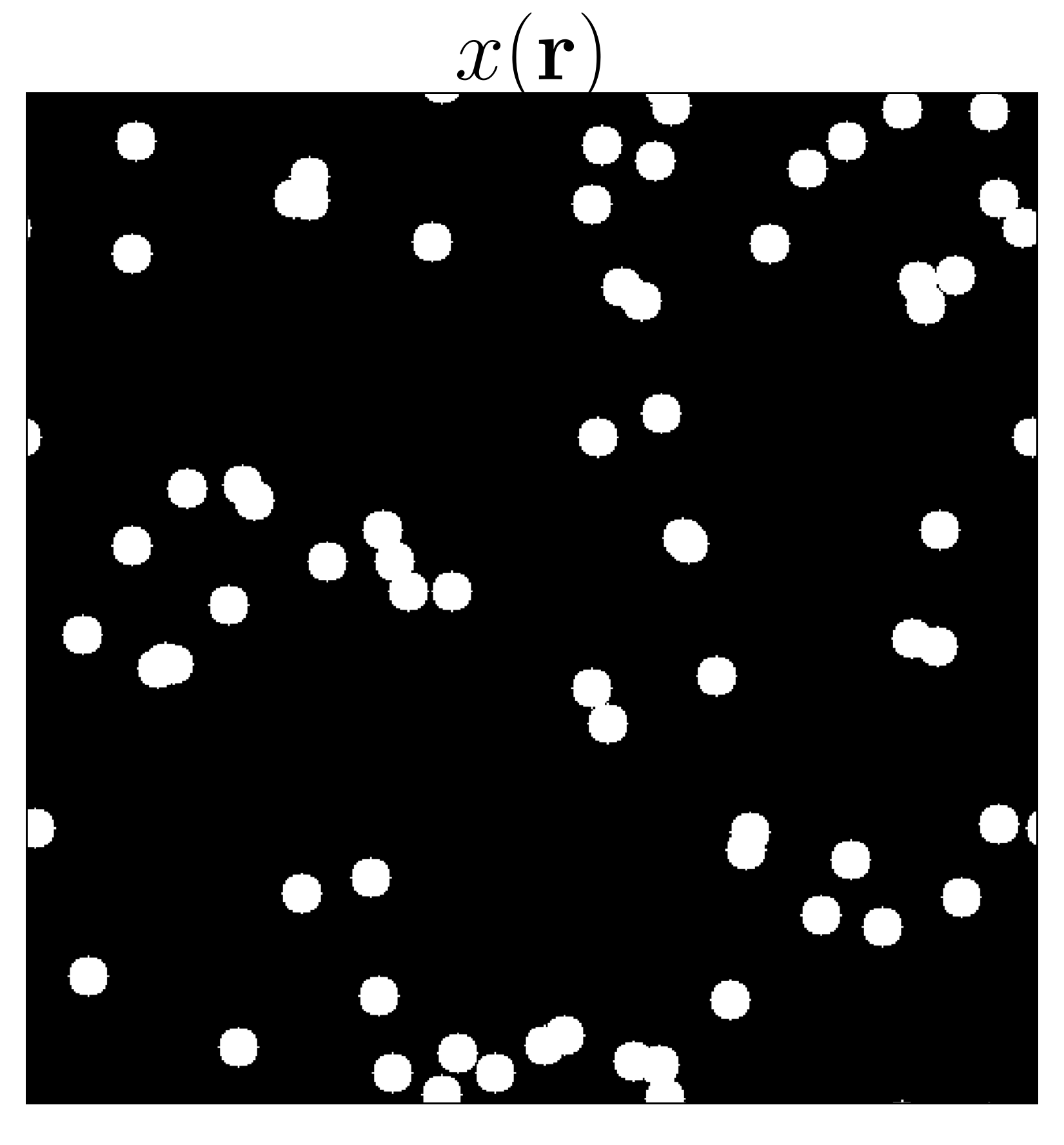}
\includegraphics[width=0.4\columnwidth]{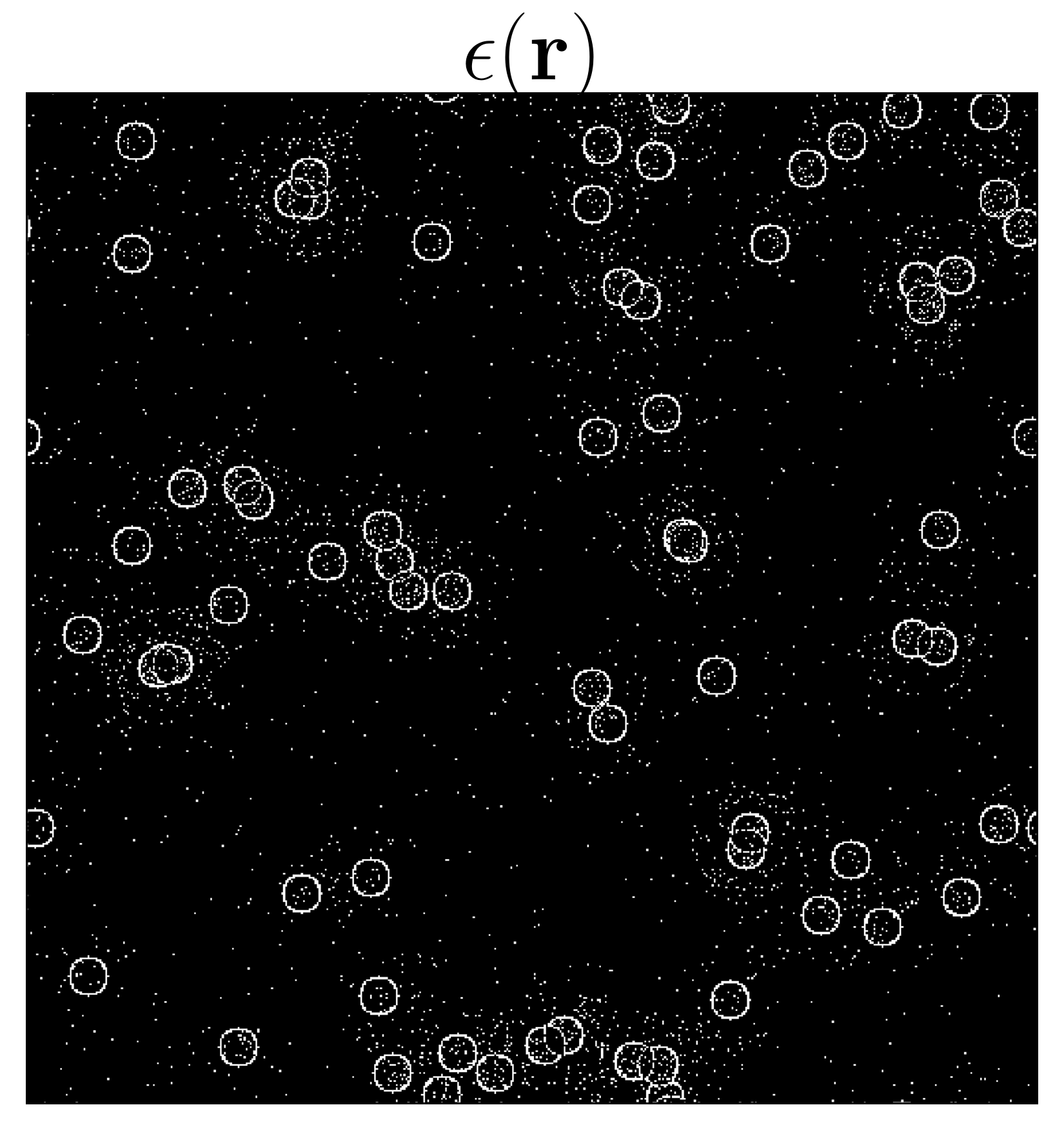}\\
\includegraphics[width=0.4\columnwidth]{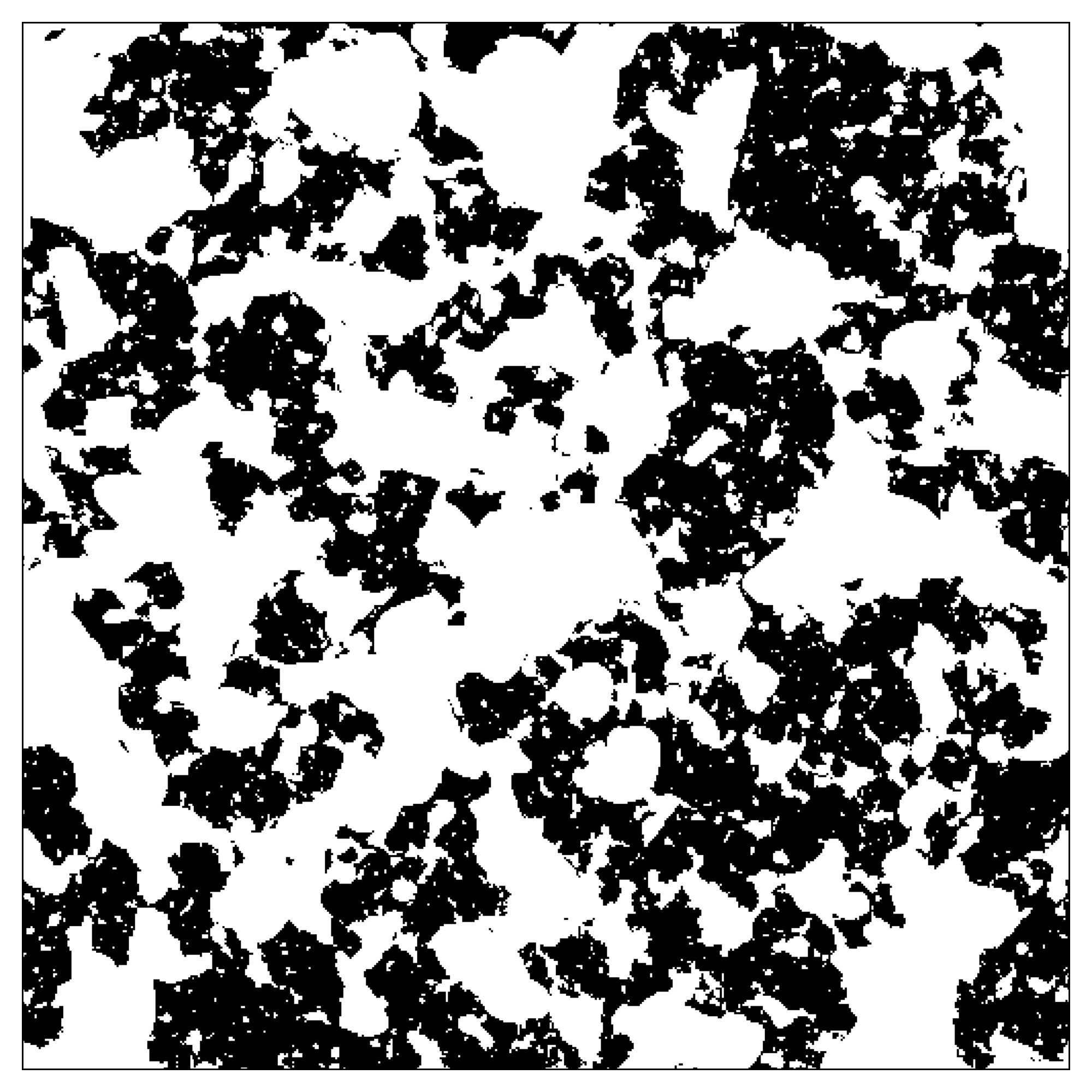}
\includegraphics[width=0.4\columnwidth]{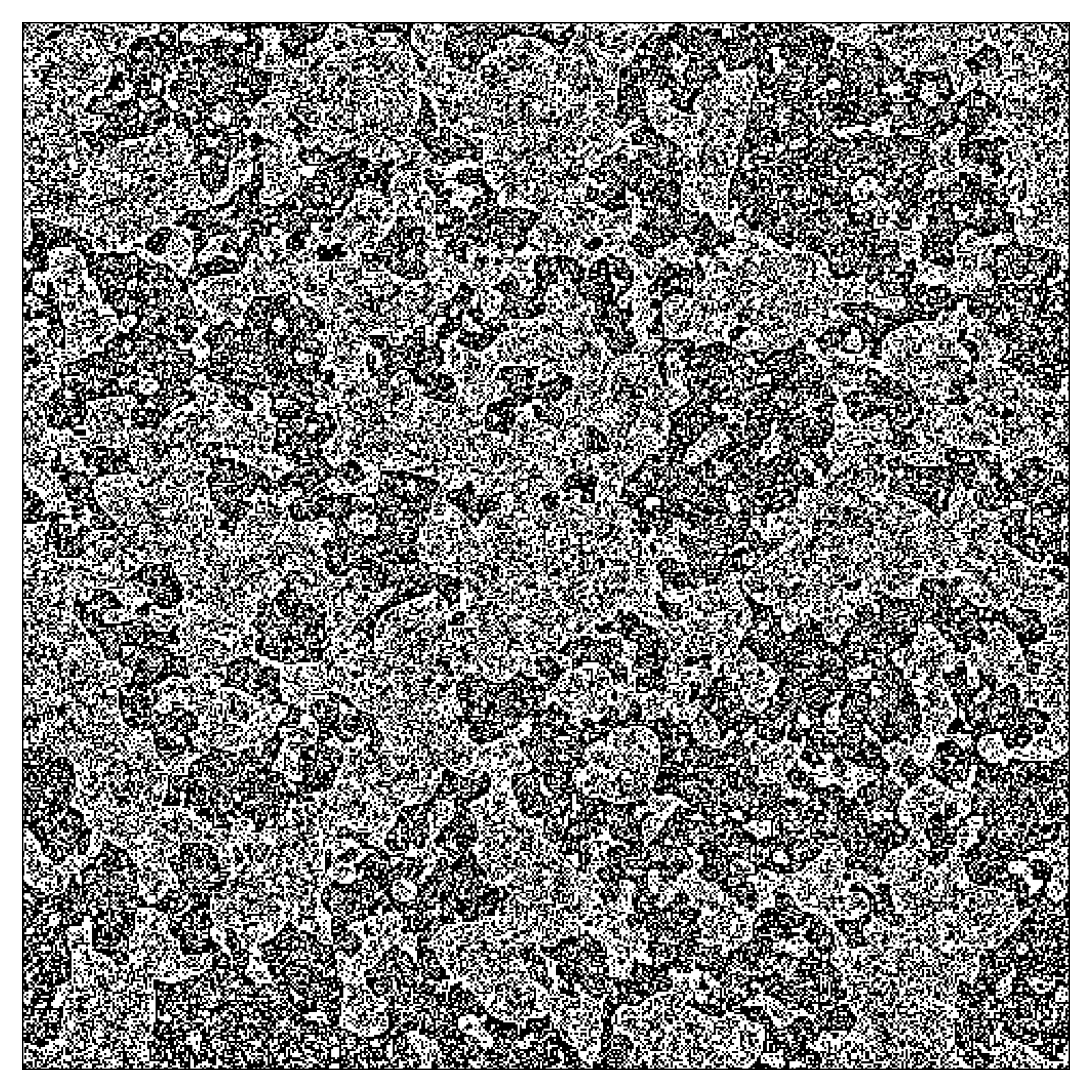}
\includegraphics[width=0.5\columnwidth]{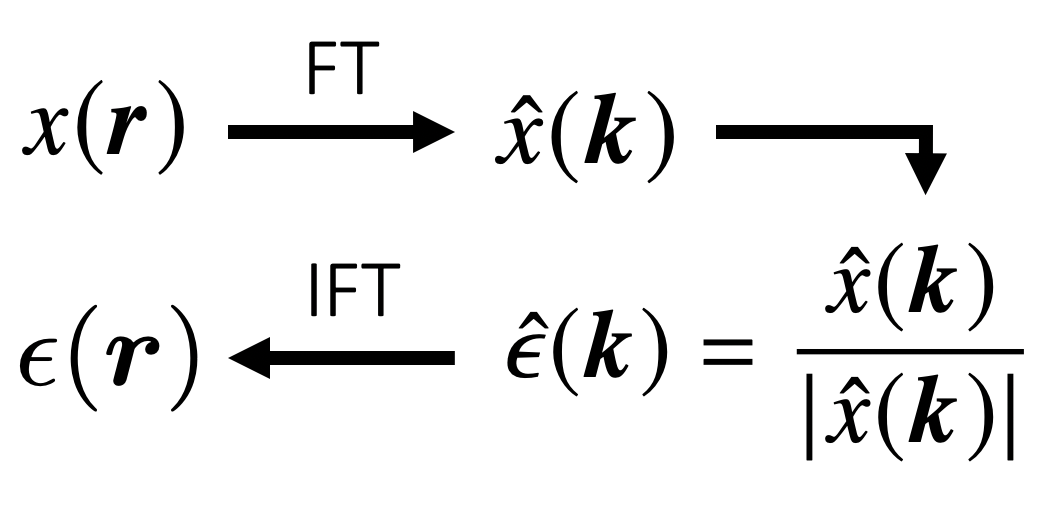}
\caption{Examples of phase information in real space. Left panels represent the ionisation field $x (\bm{r})$ and the right panel the inverse Fourier transform of the corresponding phase factor $\hat{\epsilon} (\bm{k})$. \textit{Top row:} Toy model with randomly distributed ionised bubbles. \textit{Bottom row:} Output of the 21CMFAST simulation. Both boxes have $512^2$ pixels and side length $L=400~\mathrm{Mpc}$.}
\label{fig:epsilon_vs_realbox}
\end{figure}

The top left panel of Fig. \ref{fig:epsilon_vs_realbox} shows an ionisation field made of randomly distributed ionised discs on a neutral background. The top right panel corresponds to the inverse Fourier transform of its phase factor $\hat{\epsilon}(\bm{k})$, a tracer of phase information in real space. We see that phases mostly preserve the edges of the ionised regions: they will provide information about the structure of the field. The same exercise is done in the lower panels of Fig. \ref{fig:epsilon_vs_realbox} but for an ionisation field extracted from the semi-numerical 21CMFAST simulation \citep{21cmFAST_2007} at $z=7.8$ and global ionised fraction $x_\ion{H}{II}=0.54$ (details on the simulation parameters can be found in Section \ref{sec:21cmFAST}). We see that the more complex structure of the ionisation field in this simulation is reflected in a more complex phase map. The boundaries between neutral and ionised regions are not as definite as they were for the toy model above. It will therefore be more difficult to extract information about the structure of this field from its phases.

To this day, little work has been done regarding phase information in the EoR, mostly because of the lack of a solid statistical framework to do so. \citet{thyagarajan_2018_bispectrum_phases} have shown that the 21cm signal from EoR can be detected in bispectrum phase spectra, and mention a potential use of their results for $\ion{H}{I}$ intensity mapping experiments. Most of the existing work is related to the study of galaxy clustering: although the initial conditions to matter distribution in the Universe are Gaussian, the growth of cosmic structures will lead to some non-linearity and the distributions of both amplitudes and phases will be altered \citep{watts_coles_2003,levrier_2006}. By characterising this alteration, one hopes to learn about the formation of cosmic structures and possibly about the properties of dark matter (\citealt{obreschkow_2013}, see also follow-up papers \citealt{wolstenhulme_2015,eggemeier_2015,eggemeier_2017}). We will discuss the benefits of phases compared to amplitude information in more details in Section \ref{sec:why_phases}.

\subsection{Current estimators of the size of ionised regions during reionisation}
\label{subsec:BSDs}

Characterising the morphology and topology of $\ion{H}{II}$ regions at different times through 21cm tomography is essential to the study of reionisation. For now, two types of metrics have been proposed to do so: Minkowski functionals \citep{gleser_2006_minkowski_tomography,lee_2008_topology_tomography,bag_2018_minkowski_functionals_1,chen_minkowski} and bubble size distributions (BSD). Using Minkowski functionals on tomographic images, one can learn about the topology of the reionisation process at a given time, for example the level of percolation or the shape of ionised regions. When the ionised regions largely overlap, there will likely form a unique extended percolated zone pierced by neutral tunnels \citep{bag_2018_minkowski_functionals_1,persistence_topology_2018}.
%Another interesting possibility to obtain BSD is to use the matched filter technique to optimally extract information directly from the visibilities \citep[e.g.][]{majumdar_2012_targeted_matched_filter,datta_2007_matched_filter,datta_2008_matched_filter,malloy_2013_matched_filter}, either knowing from observations at other wavelengths the location of the ionising sources -- this is what we call targeted matched filter, or without -- this is blind matched filter.
Bubble size distributions allow for more quantitative results as they give a direct estimate of the size of ionised regions. To derive BSD, three main methods can be found in the literature: the friend-of-friends algorithm (FOF), based on the connections between ionised regions \citep{iliev_2006}; the spherical average method (SPA), which looks for the largest spherical volume which can cover an ionised region and exceed a given ionisation threshold \citep{zahn_2007_SPA}; and the mean-free-path method (RMFP), based on Monte-Carlo Markov Chain (MCMC) techniques and implemented in the semi-numerical simulation 21CMFAST \citep{21cmFAST_2007,21cmFAST_2011}. The RMFP algorithm proceeds as follows: it picks a random ionised pixel, stores its distance to the closest neutral pixel in a random direction and stack these distances, measured for many pixels, in a histogram. A thorough comparison of these methods is made in \citet{lin_2016_review} for simulated clean 21cm maps and in \citet{giri_2018_bubble_sizes} for maps including observational effects. A new method, which the authors call \textit{granulometry}, has been recently presented in \citet{kakiichi_2017_HII_tomography}. The idea is to successively sieve a binary field with a spherical hole of increasing radius $R$. Depending on how many ionised pixels are sieved, one can build the probability distribution of $\ion{H}{II}$ regions sizes. In their work, the authors prove that granulometry should perform well on future SKA observations, as long as the correct observing strategies are chosen. \citet{lin_2016_review} have also extended to 3D the well-known watershed algorithm: it connects same value pixels to find isodensity lines which they interpret as the edges of characteristic structures. %Finally, \citet{giri_2018_superpixels} apply an image processing method called \textit{superpixels}, based on oversegmentation, to simulated noisy 21cm tomographic images. Their results seem more conclusive for the late stages of EoR than most other current techniques.
Note that all these methods require real space data whereas the TCF can be used directly in Fourier space.\\ %In general, it is difficult to compare the different methods used to characterise the morphology of ionised regions during reionisation because each one of them is based on a different definition of bubble size and so measures a different thing. 

We first describe in Section \ref{sec:21cm} the 21cm signal, on which the simulated observations we use are based. In Section \ref{sec:mathematical_formalism}, we develop the mathematical formalism leading to the definition of the triangle correlation function, and in particular the definitions of $n$-point correlation functions and poly-spectra. In Section \ref{sec:bubble_boxes}, we apply the TCF to simulated ionisation fields,from toy models to outputs of the 21CMFAST simulation. We relate our method to observations, and in particular to the closure phase relation, in Section \ref{sec:relation_obs}. Finally, we discuss various aspects of our results: we first give evidence for the benefits of phase information in Section \ref{sec:why_phases} and then consider computational performance in Section \ref{sec:computational_perf}. Conclusions can be found in Section \ref{sec:conclusion}.

\section{The 21cm signal}
\label{sec:21cm}

The neutral hydrogen 21cm line corresponds to the spin-flip transition of an electron between the two hyperfine levels of the ground state of the \ion{H}{I} atom. As a tracer of neutral hydrogen, it is naturally a very interesting observable to learn about the EoR. For the last decades, many efforts have been made to design experiments capable of detecting this signal, despite its low amplitude, the presence of strong foregrounds and huge calibration challenges. Seen with respect to a radio background, the CMB, the evolution of the observed differential 21cm brightness temperature writes \citep{pritchard_loeb_2012}:
\begin{equation}
\label{eq:brithness21cm}
\begin{aligned}
\delta T_\mathrm{b} & = 27\mathrm{mK} \times x_\ion{H}{I} \left( 1 + \delta_b \right)  \frac{\Omega_\mathrm{b}h^2}{0.023} \sqrt{ \frac{0.15}{\Omega_\mathrm{m}h^2}} \sqrt{ \frac{1+z}{10} } \left( 1 - \frac{T_\mathrm{CMB}}{T_\mathrm{S}} \right) \\
& \simeq 27\mathrm{mK}\ x_\ion{H}{I}  \left( 1 + \delta_b \right) \frac{\Omega_\mathrm{b}h^2}{0.023} \sqrt{ \frac{0.15}{\Omega_\mathrm{m}h^2}} \sqrt{ \frac{1+z}{10} }
\end{aligned}
\end{equation}
where $\delta_b$ is the baryon overdensity and $T_\mathrm{S}$ the spin temperature, which characterises the relative populations of the two spin states. During the late stages of reionisation, it is expected that the spin temperature will dominate the CMB temperature because of its coupling to the kinetic temperature of the gas, and we can make the approximation of the second line. %Formally, this approximation is only valid for when the Universe is more than $25\%$ ionised, but we are able to extend it to lower ionised fractions as it will not greatly impact our results. 
With this approximation, we can map the 21cm signal on the sky and, thanks to Eq. \ref{eq:brithness21cm}, interpret cold spots, i.e. regions where the signal is weaker than average, either as ionised or underdense regions; and hot spots as neutral or overdense regions. Being able to differentiate underdense from ionised regions in brightness temperature measurements is a major challenge of 21cm tomography (\citet{giri_2018_bubble_sizes}, see also \citet{giri_2018_superpixels} for an efficient way to do so). In this work however, we consider segmented data, converted into binary ionisation fields. Because of redshift, a photon emitted with a wavelength of 21cm during the EoR ($6 \lesssim z \lesssim 35$) will reach us today with a frequency $40 \lesssim \nu \lesssim 200~\mathrm{MHz}$ i.e. the frequency range that the new generation of radio interferometers, such as MWA, PAPER, LOFAR and SKA, are expected to probe.

\section{The triangle correlation function of phases}
\label{sec:mathematical_formalism}

For a real ionisation field $x(\bm{r})$ of volume $V$ in dimension $D$, the $n$-point correlation function ($n$-PCF) measures the correlations between $n$ points, described as $n-1$ vectors $\{ \bm{r}_i \} ,~i = 1, \ldots  n-1 $. It is defined as:
\begin{equation}
\label{eq:def_n-PCF}
\Xi_n (\bm{r}_1, ..., \bm{r}_{n-1}) = \frac{1}{V} \int_V \mathrm{d}^D r \, \prod_{j=1}^{n-1} x (\bm{r} + \bm{r}_j ),
\end{equation}
where $\bm{r}_n=0$. If we average this definition over all possible rotations, we obtain the isotropic $n$-PCF $\xi_n$. As can be seen in Eq. \ref{eq:def_n-PCF}, the $n$-PCFs are convolutions and it will be easier to compute them in Fourier space. The Fourier transform of the $n$-PCF is called the $n^\mathrm{th}$ poly-spectrum $P_n$ such that
\begin{equation}
P_n \left( \bm{k}_1, \ldots \bm{k}_{n-1} \right) = \hat{x} (\bm{k}_1) \ldots \hat{x} (\bm{k}_{n-1})\, \hat{x} \left( -\Sigma \bm{k}_j \right)
\end{equation}
and 
\begin{equation}
\begin{aligned}
\Xi_n \left( \bm{r}_1 , \ldots \bm{r}_{n-1} \right) = \left[ \frac{V}{(2\pi)^D} \right] ^{n-1} \int \mathrm{d}^D k_1 \, \mathrm{e}^{i \bm{k}_1 \cdot \bm{r}_1} \, \ldots \\ \times \int \mathrm{d}^D k_{n-1} \, \mathrm{e}^{i \bm{k}_{n-1} \cdot \bm{r}_{n-1}} \ P_n \left( \bm{k}_1, \ldots \bm{k}_{n-1} \right).
\end{aligned}
\end{equation}
For $n=2$, $P_2$ is called the power spectrum $\mathcal{P}(\bm{k})$ and for $n=3$, we have the bispectrum $\mathcal{B}(\bm{k}, \bm{q})$ such that
\begin{subequations}
\begin{align}
& \mathcal{P} (\bm{k}) = \hat{x} (\bm{k})\, \hat{x} (- \bm{k}) = \vert \hat{x} (\bm{k}) \vert ^2, \label{eq:def_power_spectrum} \\
& \mathcal{B} (\bm{k}, \bm{q}) = \hat{x} (\bm{k})\, \hat{x} (\bm{q})\, \hat{x} (- \bm{k} - \bm{q}),
\label{eq:def_bspectrum}
\end{align}
\end{subequations}
where in the first equation we have used the fact that, because $x \left( \bm{r} \right)$ is a real field, its Fourier transform verifies $\hat{x} (-\bm{k}) = \hat{x}^{*} (\bm{k})$. Note that we can also write the bispectrum in terms of three $k$-vectors $(\bm{k}_1, \bm{k}_2, \bm{k}_3)$ forming a closed triangle i.e $\bm{k}_1 + \bm{k}_2 + \bm{k}_3 = 0$.

Consider the 3-point correlation function (3-PCF), i.e. the inverse Fourier transform of the bispectrum:
\begin{equation}
\label{eq:xi_def}
\Xi_3 \left(\bm{r}, \bm{s} \right) = \frac{V^2}{(2\pi)^{2D}} \iint \mathrm{d}^D k \ \mathrm{d}^D q \ \mathrm{e}^{i(\bm{k}\cdot \bm{r} + \bm{q} \cdot \bm{s})} \, \mathcal{B}(\bm{k}, \bm{q}).
\end{equation}
To study filamentary structures in matter fields, \citet[][hereafter \citetalias{obreschkow_2013}]{obreschkow_2013} consider the 3-PCF in Eq. \ref{eq:xi_def} for $\bm{r}=-\bm{s}$ i.e. three equidistant points forming a straight line. Because here we look for spherical structures, we will consider two vectors $\bm{r}$ and $\bm{s}$ forming an equilateral triangle, as it is the three-point shape closest to a sphere.
In this case, $\bm{s}$ is just $\bm{r}$ rotated by an angle $\pi/3$:
\begin{equation}
\left\{ \begin{aligned}
& s_x = \frac{1}{2} r_x - \frac{\sqrt{3}}{2} r_y \\
& s_y = \frac{\sqrt{3}}{2} r_x + \frac{1}{2} r_y \\
& s_z = r_z\\
\end{aligned}
\right.
\end{equation}
where the 2D case is limited to the first two equations\footnote{Note that in the actual computation of the triangle correlation function, we will consider rotations not only around the $z$-axis but also around the other two.}. Let 
\begin{equation}
\bm{p} = \begin{pmatrix}
    k_x + \frac{1}{2} q_x + \frac{\sqrt{3}}{2} q_y   \\
    k_y -  \frac{\sqrt{3}}{2} q_x + \frac{1}{2} q_y \\
    k_z + q_z
\end{pmatrix},
\end{equation}
then we can write $ \bm{k} \cdot \bm{r} + \bm{q} \cdot \bm{s} = \bm{p} \cdot \bm{r} $ in Eq. \ref{eq:xi_def}.
If we choose to consider the phase factor of the bispectrum rather than its full form (see Eq. \ref{eq:def_phase_factor}), we find the modified 3-PCF in dimension $D$
\begin{equation}
\Xi^{*}_3 (\bm{r}) = \frac{V^2}{(2 \pi)^{2D}} \iint \mathrm{d}^D k \, \mathrm{d}^D q \ \mathrm{e}^{i \bm{p} \cdot \bm{r}} \, \frac{\mathcal{B} \left( \bm{k}, \bm{q} \right)}{\vert \mathcal{B} \left( \bm{k}, \bm{q} \right) \vert}.
\end{equation}
Note that, according to \citet{eggemeier_2015}, the phase factor of the bispectrum can be directly related to the normalised bispectrum defined in \citet{bispectrum_catherine_2017} and later used in \citet{bispectrum_xray_2019} and \citet{trott_watkinson_2019}. Assuming ergodicity, a rotational average of the above expression gives the isotropic modified 3-PCF:
\begin{equation}
\xi^{*}_3(r) \equiv  \frac{V^2}{(2 \pi)^{2D}} \iint  \mathrm{d}^D k \ \mathrm{d}^D q \ \omega_D \left(pr\right)  \, \frac{\mathcal{B} \left( \bm{k}, \bm{q} \right)}{\vert \mathcal{B} \left( \bm{k}, \bm{q} \right) \vert},
\end{equation}
where $\omega_D$ is the window function defined by
\begin{equation}
\omega_D(x) = 
\left\{ \begin{aligned}
    & \frac{\mathrm{sin} \left(x \right)}{x}  & \mathrm{if} \ D=3,  \\
    & J_0(x) & \mathrm{if} \ D=2,\\
    \end{aligned}
    \right.
\end{equation}
for $J_0(x)$ the Bessel function of the first kind and order 0.
Numerically, we will need to discretise the integral as
\begin{equation}
\label{eq:def_iso_modified_3PCF}
\xi^{*}_3(r) =  \sum_{\bm{k}} \sum_{\bm{q}} \omega_D\left(pr\right)  \, \frac{\mathcal{B} \left( \bm{k}, \bm{q} \right)}{\vert \mathcal{B} \left( \bm{k}, \bm{q} \right) \vert},
\end{equation}
for a periodic box with physical side length $L$, divided into $N^D$ cells. Each cell has a width $\Delta x = L/N$. In Fourier space, this box transforms into a box of same dimensions ($N^D$) but of side length $ 2\pi N /L$ and spacing $\Delta k = 2\pi/L$. The largest mode $k = 2\pi/\Delta x$ has the smallest wavelength. If we assume phases are uncorrelated below a given scale, then the modes whose wavelengths are smaller than this scale will have random phases. When the resolution is improved, i.e. $\Delta x$ reduced, the number of such modes increases and random phase terms are added to Eq. \ref{eq:def_iso_modified_3PCF} so that the signal eventually diverges. Following \citetalias{obreschkow_2013}, we introduce a cut-off $k \leq \pi / r$ on the sums of Eq. \ref{eq:def_iso_modified_3PCF} to limit this number. Similarly, when increasing the size of the box $L$, we add modes with wavelength larger than the largest correlation scale within the box, and therefore add random phase terms to the sum. Because $\Delta k = 2\pi / L$, the number of modes scales as $L^D$ and $s(\bm{r})$ diverges as $L^{3D/2}$. As discussed in \citetalias{obreschkow_2013}, we introduce the pre-factor $(r/L)^{3D/2}$ to Eq. \ref{eq:def_iso_modified_3PCF} to remove this divergence. These changes applied to the isotropic modified 3-PCF $\xi^{*}_3(r)$ define the triangle correlation function of phases:
\begin{equation}
\label{eq:def_triangle_correlation_function}
s (r) = \left( \frac{r}{L} \right) ^{3D/2} \sum_{k,q \, \leq \, \pi/r} \omega_D \left(pr\right) \, \frac{\mathcal{B} \left( \bm{k}, \bm{q} \right)}{\vert \mathcal{B} \left( \bm{k}, \bm{q} \right) \vert}.
\end{equation}
The imaginary part of $s(\bm{r})$ is zero therefore we only consider its real part in the following sections. 

All these derivations have been done with $x(\bm{r})$ being the ionisation field $x_\ion{H}{II}$. If we consider the neutral field $x_\ion{H}{I}$, we have $x_\ion{H}{II} = 1 - x_\ion{H}{I}$ and $\hat{x}_\ion{H}{II} (\bm{k}) = - \hat{x}_\ion{H}{I}(\bm{k})$ which we can plug back into the equations above to find $s_\ion{H}{II}(r)= -s_\ion{H}{I}(r)$. Therefore when applied to a mostly ionised field containing a few remote neutral islands, the correlations dominating the signal will be related to $\ion{H}{I}$ regions and the signal will be negative. In particular, because 21cm interferometric images trace the neutral gas in the sky, if we apply our method on 21cm brightness temperature maps, we obtain the exact same signal as for the corresponding ionisation field, but with a reversed sign. This also implies that during the middle stages of the reionisation process, positive and negative (respectively, $\ion{H}{II}$ and $\ion{H}{I}$) correlations overlap, and the TCF flattens down.

\section{Application to simulated ionisation fields}
\label{sec:bubble_boxes}

We generate boxes of $N^2$ cells filled with randomly distributed ionised disks. Although the derivations performed in Section \ref{sec:mathematical_formalism} are valid for a two or three-dimensional ionisation field, we now limit our work to 2D boxes, as they are closer to what one would observe with a radiotelescope. These boxes are binary: an ionised region will have pixels of value 1, whereas a neutral zone will be filled by zero pixels. Because the UV photons emitted by early galaxies have a very short mean free path in the surrounding neutral IGM, the boundary between ionised and neutral regions is expected to be sharp, and such a binary model is acceptable \citep{furlanetto_2006_HIIregions}. Each simulated box has periodic boundary conditions and ionised regions are allowed to overlap\footnote{When they overlap, the maximum pixel value is set to 1, since it is a proxy for ionised level.}. In a more realistic field, we would expect the ionised bubbles to be clustered around overdense regions rather than randomly distributed but we choose to first ignore this effect, as we will later consider it when applying our method to the 21CMFAST simulation. For each box, we compute the triangle correlation function as defined in Eq. \ref{eq:def_triangle_correlation_function} for a range of correlation scales $r$. We compare the resulting plot with the known size of the ionised bubbles filling the box. The programme we developed to compute these correlations is publicly available online\footnote{\url{https://github.com/adeliegorce/Triangle_correlations}}.

\subsection{Picking up characteristic scales}
\label{subsec:scales_random}

\begin{figure}
	\centering
	\includegraphics[width=0.8\columnwidth]{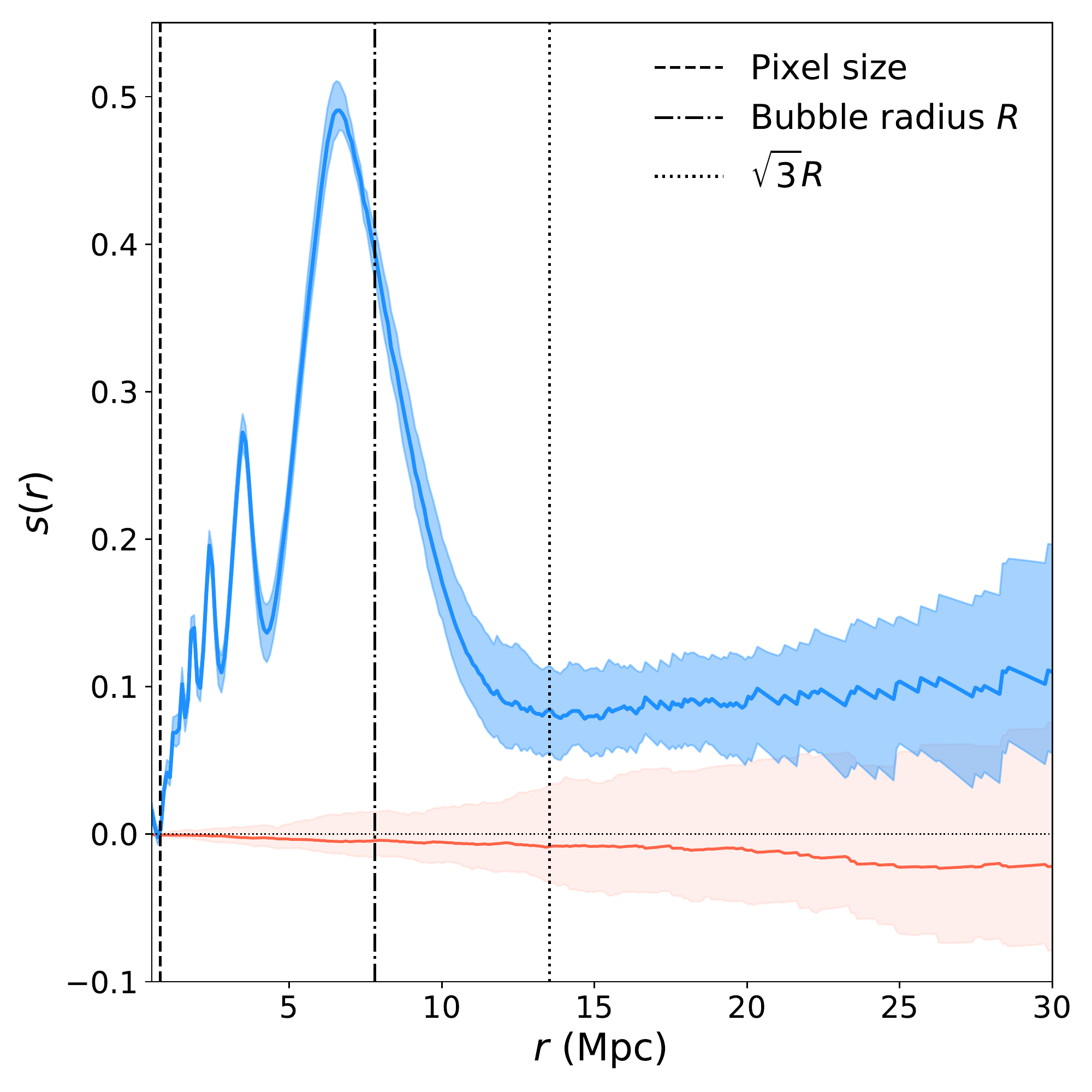}
    \caption{Triangle correlation function for a box of $512^2$ pixels and side length $L=400~\mathrm{Mpc}$ filled with 70 binary bubbles of radius $R=10$ (pixel units). The shaded area corresponds to the $95\%$ confidence interval as the function was computed for 30 different realisations of the same box. The triangle correlations of 30 Gaussian random fields of same dimensions are represented as the red line (mean value) and matching shaded area ($95\%$ confidence interval).}
    \label{fig:R10_70_binary}
\end{figure}

Fig. \ref{fig:R10_70_binary} shows the TCF computed for 30 realisations of a 2D box of $512^2$ pixels and side length $L=400~\mathrm{Mpc}$ filled with 70 binary bubbles of radius $R=10$ (in pixel units) i.e. about $7.8~\mathrm{Mpc}$. The solid line is the mean value for all 30 boxes, and the shaded area represents the $95\%$ confidence interval. Although the signal is very similar for all realisations of the box at small correlation scales, there is more variance at larger scales, as $r$ gets closer to $L$. To see if this scattering is a numerical or physical effect, we compute the triangle correlations of 30 boxes of same dimensions, but with random Fourier phases; the result is added in red to the figure. In this case, the mean signal is close to zero but variance can still be seen at large scales. Interestingly, \citet{eggemeier_2017} note the same scatter in the line correlation function $\ell (\bm{r})$ of density fields. The line correlation function is the 3-PCF of Fourier phases for three equidistant points aligned in real space \citep{obreschkow_2013}. Through its analytical derivation, they trace this scatter back to the Gaussian part of the covariance matrix of $\ell (\bm{r})$. From another point of view, we see that the mean signal flattens out when many realisations of the same box are considered. This suggests that the power seen on large scales corresponds to correlations between three points located in separate bubbles and so traces a form of correlation in the bubble locations. Because our simulation randomly distributes bubbles in the box, it is then only natural that these correlations would average out. The remaining non-zero value on large scales decreases as we increase the number of bubbles in the box and therefore can be assimilated to shot noise. Note that the oscillations on the left side of the peak are due to the sharp edges of the bubbles: when symmetrical Gaussian distributions are used instead of binary bubbles, or once the field is smoothed by a given angular resolution, they vanish. We refer the interested reader to \citet{bispectrum_catherine_2017} and \citet{bispectrum_suman_2018}, where the authors discuss the various problems induced by solid spheres in the bispectrum. Finally, a comparison of numerical results with an analytical derivation of the TCF for a similar toy model can eb found in Appendix \ref{appendix:derivation}.

\begin{figure}
\centering
	\includegraphics[width=0.85\columnwidth]{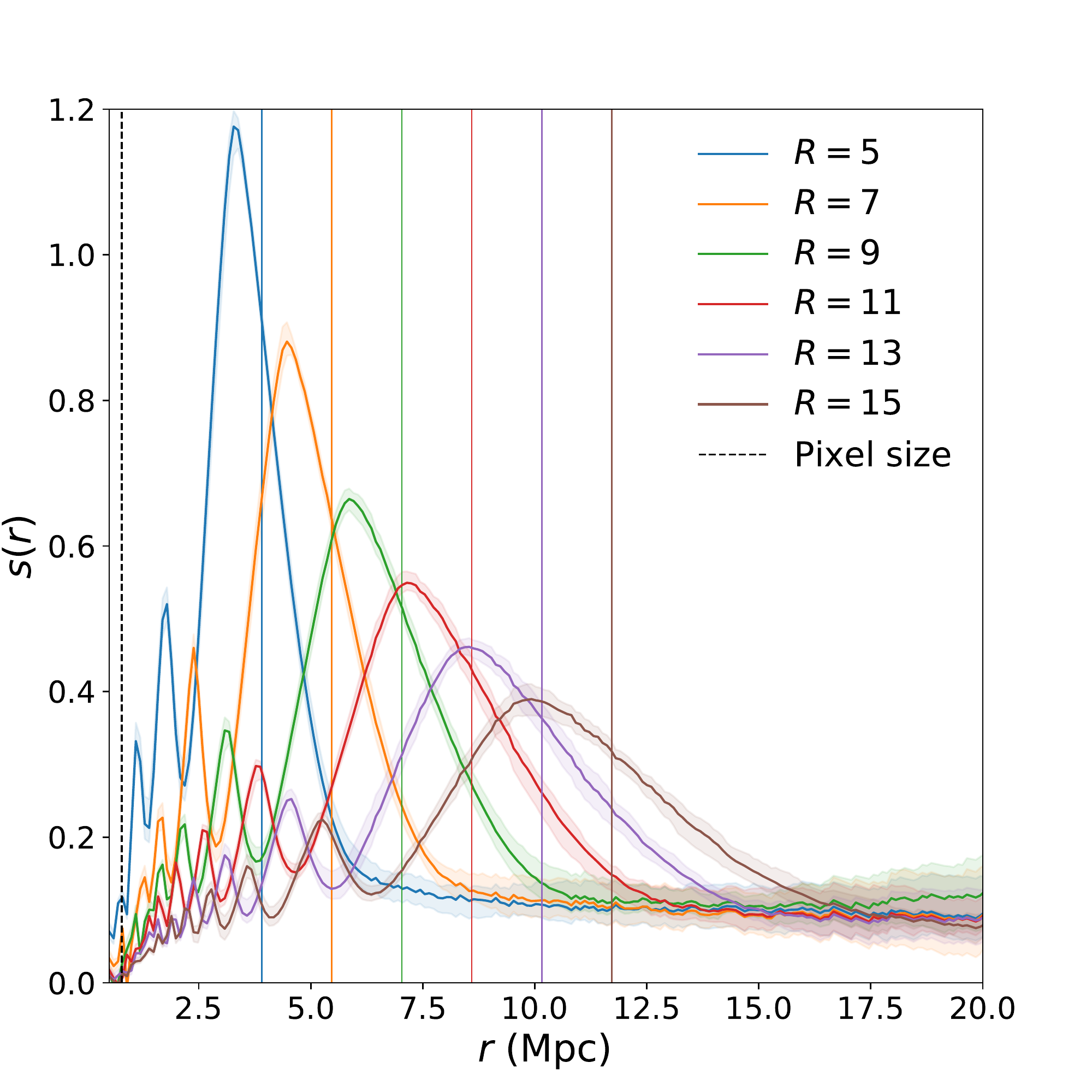}
    \includegraphics[width=0.85\columnwidth]{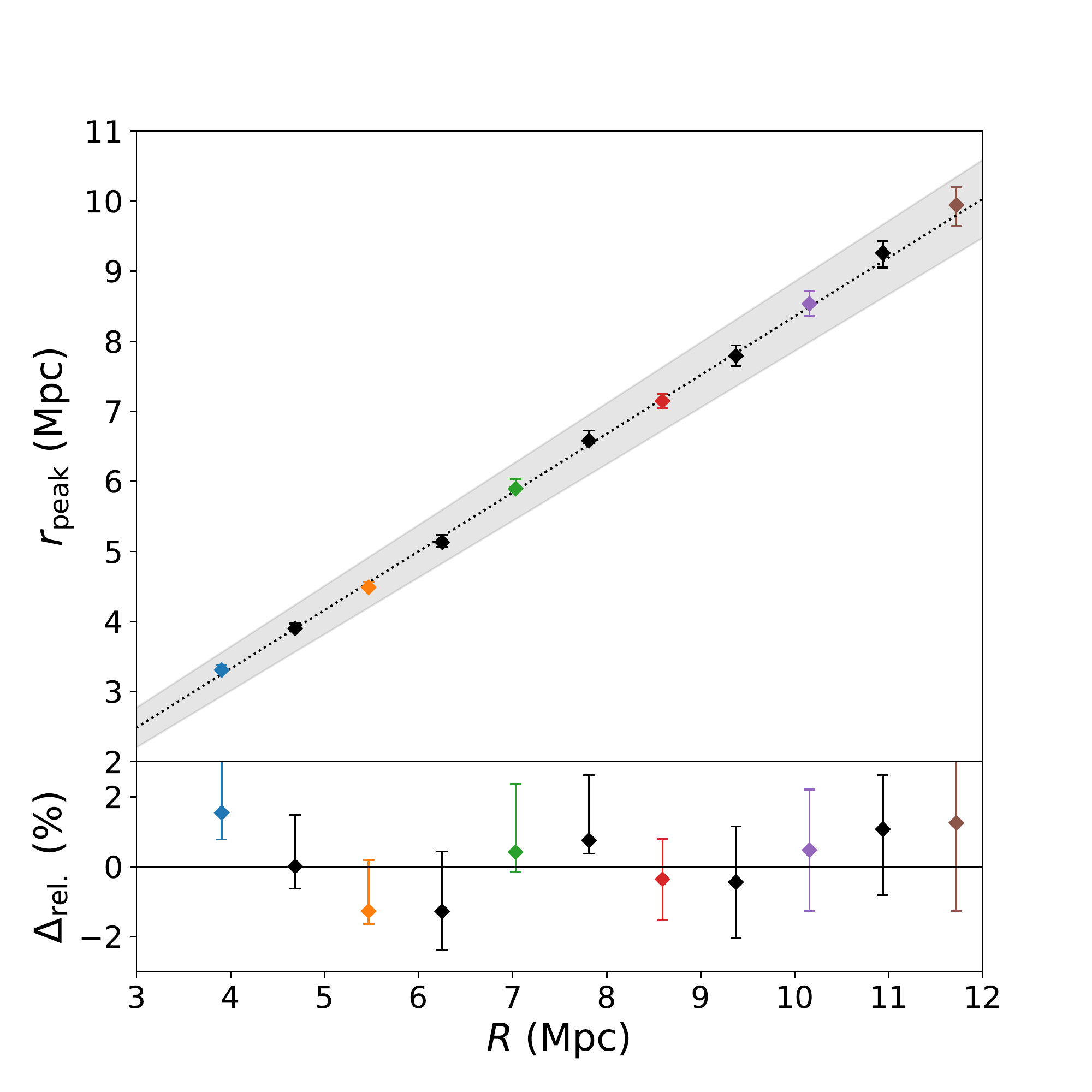}
    \caption{\textit{Upper panel:} Triangle correlation function for 2D boxes 512 pixels, with box side length $L=400~\mathrm{Mpc}$ and filled with 50 binary bubbles of different radii (see legend, in pixel units). Vertical lines are $R$ for the box of the corresponding colour. \textit{Lower panel:} Relation between the scale at which the triangle correlation function peaks $r_\mathrm{peak}$ and the size of the bubbles in the corresponding box $R$. The dotted line the maximum likelihood linear relation between the two with $95\%$ confidence interval as the shaded area. The lower panel gives the relative distance to this result for each point.}
    \label{fig:R_50_binary}
\end{figure}

The triangle correlation function probes equilateral triangles inscribed in the ionised discs of the box, and simple geometry shows that these have a side length of $\sqrt{3}R$, a scale at which we would expect $s(r)$ to peak. On Fig. \ref{fig:R10_70_binary}, there is a clear peak in the signal, but at $r \simeq 6.7 ~\mathrm{Mpc} < \sqrt{3}R$. To check the physical meaning of this peak, we generate many boxes of same dimensions and fill each one of them with 50 binary bubbles of a given radius. Results are gathered in the upper panel of Fig. \ref{fig:R_50_binary}: the TCF peak shifts to larger scales as bubbles increase in size. We plot the peaking scale $r_\mathrm{peak}$ as a function of the bubble radius $R$ in the lower panel of Fig. \ref{fig:R_50_binary}, where the error bars correspond to the $95\%$ confidence interval for 10 realisations of the same box. %Note that because all the boxes considered have the same dimensions, sample variance make the error bars larger for larger bubbles. 
Points are closely aligned, and a MCMC fit to a linear relation gives the following results, where both $r_\mathrm{peak}$ and $R$ are in $\mathrm{Mpc}$ and the uncertainty corresponds to the standard deviation on sampled parameters:
\begin{equation}
\label{eq:rvsR_binary}
r_\mathrm{peak} = \left( 0.838 \pm 0.015 \right) \, R + \left( - 0.028 \pm 0.099 \right).
\end{equation}
We trace this shift between the peaking scale and the expected peak ($\sqrt{3}R$) back to the window function: when we take $\omega_D(x)=1$ in Eq. \ref{eq:def_triangle_correlation_function} to compute the TCF, the signal has a shape similar to what is found for the correct window function, but is stretched over the $x$-axis, so that it peaks at a scale $R \leq r_\mathrm{peak} \leq \sqrt{3}R$.

Here, we chose to work with a binary field, with no partially ionised regions where $0< x_\mathrm{H_{II}} <1$. To see if our results hold when such regions exist, as there would be if X-ray photons significantly contributed to reionisation \citep{visbal_loeb_2012} or if some observed regions are unresolved, we perform the same test but for symmetric 2D Gaussian distributions centred on the bubble radius $R$. The box dimensions are the same as before. As for binary bubbles, we observe a peak in the triangle correlation function at scales slightly smaller than $\sqrt{3}R$. A linear relation can still well describe the correlation between $r_\mathrm{peak}$ and $R$ but there is more scatter around the model. This is mostly due to the peaks being difficult to locate: on average, the signal is much flatter for Gaussian than for binary bubbles. 

 Note that we checked that these results are independent of the number of bubbles in the box. Similarly to what \citetalias{obreschkow_2013} find, the number of bubbles (or filling fraction) will only impact the amplitude of the signal, not its shape. Indeed, for a given radius, the amplitude of the signal will decrease as the number of bubbles -- and so the filling fraction of the box, increases. This is likely due to the fact that increasing the number of bubbles increases the number of different translations between objects within the box and so randomises the phase terms \citep{eggemeier_2015}. In general, we find that as long as the filling fraction does not exceed $60\%$, there is still a clear peak in the signal and we can infer a characteristic scale. This corresponds to the fact that in the late stages of the reionisation process, the ionised regions have largely overlapped and they have no characteristic shape and size anymore. We can compare these results to those of \citet{bharadwaj_2005} who, through a more observational approach, relate the bispectrum of $\ion{H}{I}$ fluctuations to the correlations between the visibilities measured at three different baselines of an interferometer. They generate the same kind of toy models as us and find that their signal increases with the size of the ionised regions, but overlaps prevent their method from being used in the mid-stages of reionisation ($x_\ion{H}{II} \geq 0.5$). %Their approach is quite similar to ours as they only consider combinations of baselines forming an equilateral triangle; however, it requires the bubble radius size as an input.

\begin{figure}
	\centering
	\includegraphics[width=0.8\columnwidth]{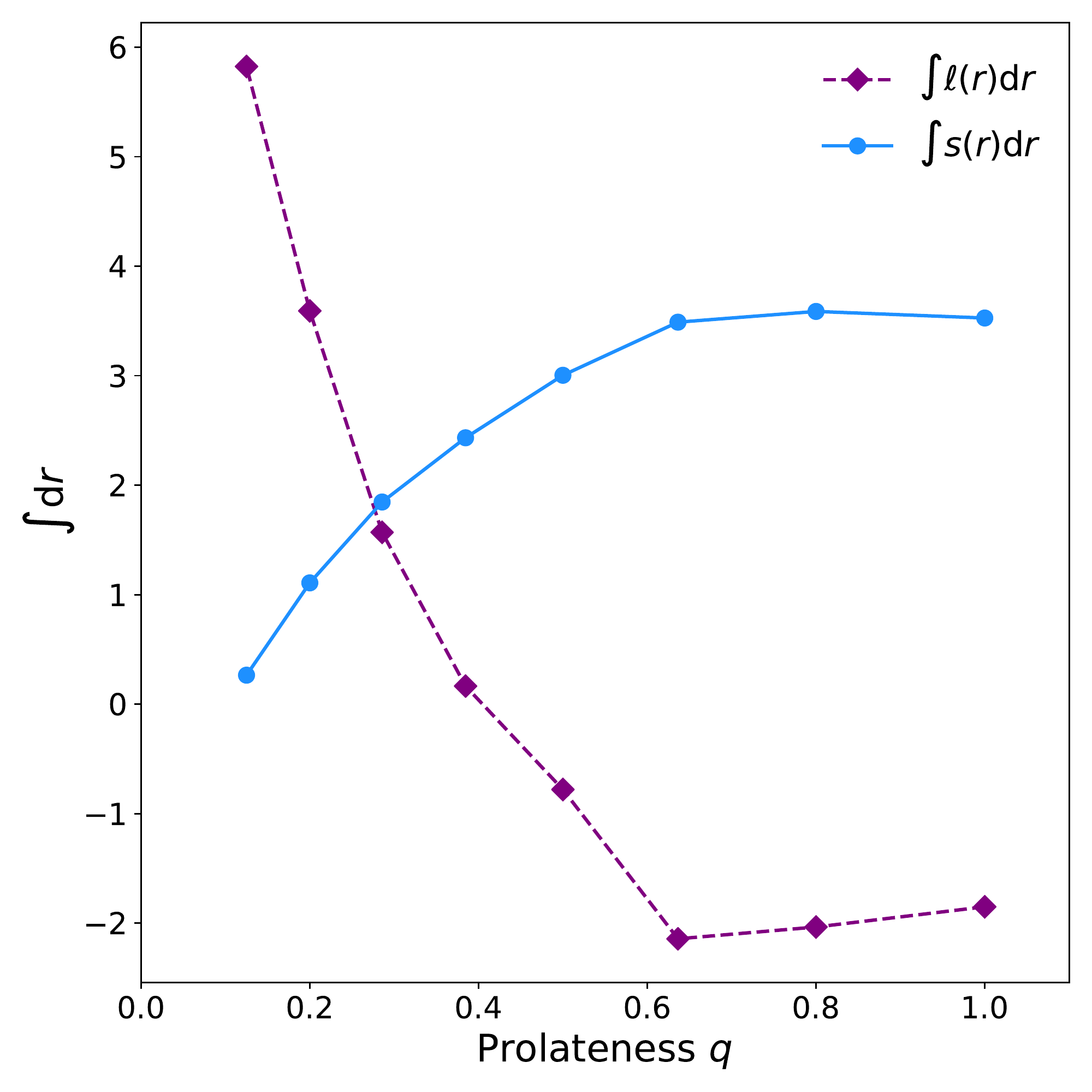}
    \includegraphics[width=0.95\columnwidth]{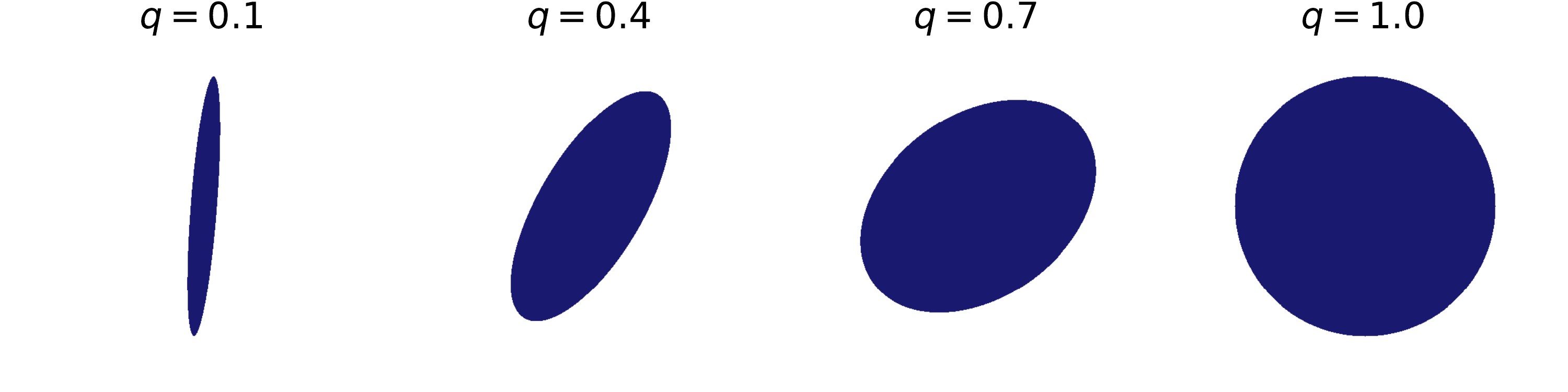}
    \caption{Evolution of the integral of the line (dashed purple) and triangle (solid blue) correlation functions with the prolateness of shapes filling up the considered box. Dimensions of studied boxes are again $N=512$ and $L=400~\mathrm{Mpc}$.}
    \label{fig:q_vs_int_s_l}
\end{figure}

\subsection{Further tests}
\label{subsec:toy_model_properties}

\begin{figure*}
\centering
    \subfloat[]{\includegraphics[height=2.1in]{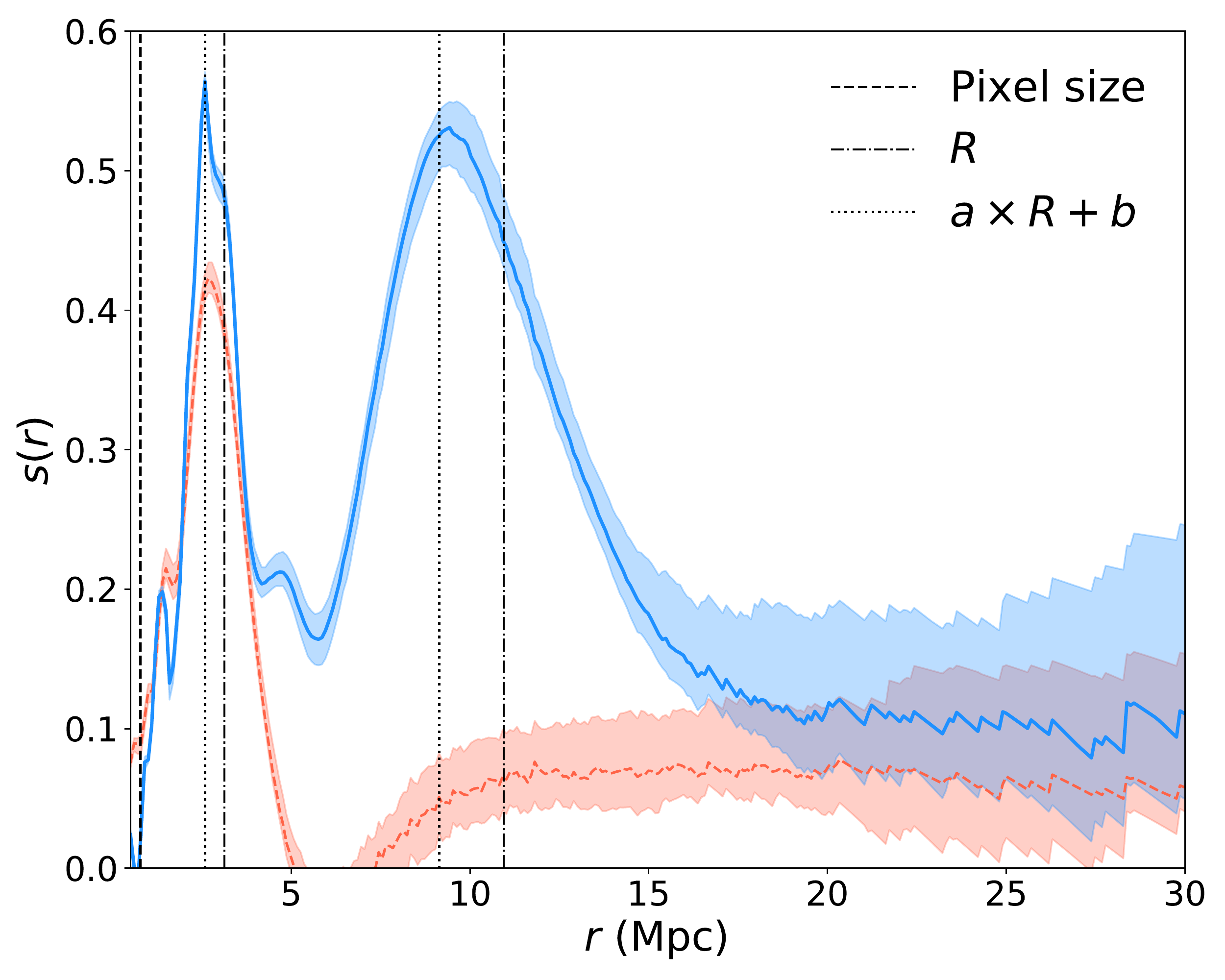}\label{fig:2sizes_binary}}
    \hspace{0.3in}
    \subfloat[]{\includegraphics[height=2.1in]{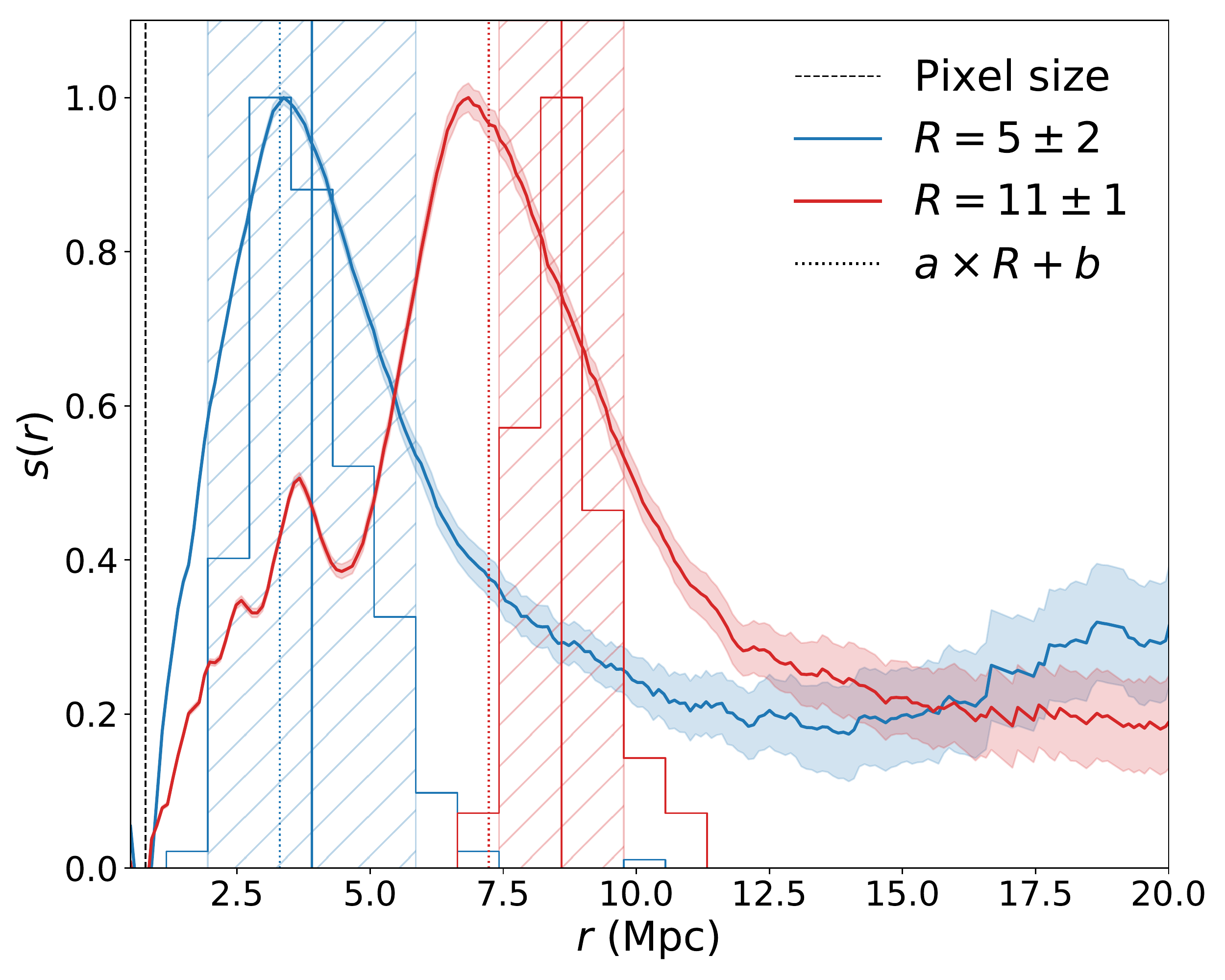}\label{fig:radius_distri}}
\caption{\textit{(a)} Triangle correlation function for 10 realisations of a box filled with 20 binary bubbles of radius 14 and 60 of radius 4 (in pixel units). Box dimensions are $N=512^2$ and $L=400\, \mathrm{Mpc}$. Vertical dashed-dotted lines correspond to the two bubble radii, the vertical dotted lines to the radii scaled by the linear relation from Eq. \ref{eq:rvsR_binary}. Triangle correlations for 60 binary ellipses of the same dimensions are represented by the orange dashed line. \textit{(b)} Triangle correlation function for 10 realisations of boxes filled with binary bubbles whose radii are selected from a log-normal distribution to reach a filling fraction of $\bar{x}_\ion{H}{II} = 0.09$. Box dimensions are $N=512^2$ and $L=400\, \mathrm{Mpc}$. The TCF (scaled to 1, solid lines) is compared to the actual distribution of radii, extracted directly from the simulation and shown as a histogram. Hatched areas correspond to $R \pm \sigma$. Vertical solid (resp. dotted) lines correspond to the radius (resp. radius scaled by the linear relation from Eq. \ref{eq:rvsR_binary}).}
\end{figure*}

The TCF seems to correctly pick up the characteristic scale of spherical regions in an ionisation field. Because we want to use this method directly on Fourier data, for which no real space image will be available, we need to ensure that the peaking scale corresponds to spherical regions only. In their work, \citetalias{obreschkow_2013} define the line correlation function (LCF), which is analogous to our triangle correlation function but for two vectors $\bm{r}$ and $\bm{s}$ aligned ($\bm{s} = - \bm{r}$). Therefore $\ell (\bm{r})$ probes elongated rather than spherical structures and writes:
\begin{equation}
\label{eq:def_line_correlation_function}
\ell (r) = \left( \frac{r}{L} \right) ^{3D/2} \sum_{k,q \, \leq \, \pi/r} \omega_D \left(  \vert\bm{k} - \bm{q} \vert r \right) \, \frac{\mathcal{B} \left( \bm{k}, \bm{q} \right)}{\vert \mathcal{B} \left( \bm{k}, \bm{q} \right) \vert},
\end{equation}
where the window function has not changed compared to our expression but is now a function of $\vert\bm{k} - \bm{q} \vert r$.
\citetalias{obreschkow_2013} find the LCF to be close to flat when there are only spherical structures in the field considered (see their Figs. 3 and 4). To check the significance of each correlation function, we compute both of them for fields filled with bubbles of various sizes and compare the results. We find that the LCF also picks up the bubble size, with a peak located at roughly the same correlation scales as for triangle correlations, but with a much weaker amplitude. 
To push the comparison further, we compute the TCF and the LCF for 2D boxes filled with more or less stretched binary ellipses. We define the prolateness $q$ of an ellipse as the ratio of its semi-minor axis to its semi-major axis so that an ellipse with prolateness $q=1$ is a flat disk. Fig. \ref{fig:q_vs_int_s_l} shows the evolution of the integral of the two correlation functions as a function of the prolateness of the objects in the box \footnote{Note that we integrate only on the correlation range $0.5~\mathrm{Mpc} \leq r \leq 20~\mathrm{Mpc}$ where the signal is more reliable (see Sec. \ref{subsec:scales_random}).}. As $q \rightarrow 1$, the triangle correlations signal increases, while the line correlations signal decreases, even reaching negative values for $q>0.4$. On the contrary, for $q=0.1$, the TCF is close to zero and the LCF shows very strong signal. %Note that triangle correlations overshoot line correlations as soon as $q=0.25$ i.e. when the ellipsoid is still very stretched. 
This confirms that the TCF mostly picks up spherical structures and therefore can safely be used to infer characteristic bubble sizes directly from Fourier data.

In reality, the reionisation process will not be as homogeneous as our toy models: it is unlikely that all ionised bubbles should have the same radius at a given redshift. It would therefore be useful if the triangle correlation function could differentiate between bubbles of different radii. To test for this, we generate a box of $512^2$ pixels and side length $L=400~\mathrm{Mpc}$, filled with 20 binary disks of radius $14~\mathrm{px}$ and 60 of radius $4~\mathrm{px}$. Results are displayed on Fig. \ref{fig:2sizes_binary} for 20 realisations of this box. We can clearly distinguish two peaks, which seem to match the bubble radii (dash-dotted vertical lines), once scaled by the coefficients of Eq. \ref{eq:rvsR_binary} (dotted vertical lines). Note that this result holds when we replace the binary bubbles by Gaussian disks. To ensure that the scales picked up by our function correspond to two different sizes of bubbles and not to the length and width of elongated structures, we compute the triangle correlations of a box filled with 60 binary ellipses of the same dimensions as the bubbles above i.e. a semi-major axis $\alpha=14~\mathrm{px}$ and a semi-minor axis $\beta=4~\mathrm{px}$. The resulting TCF, shown as an orange dashed line on the figure, differs largely from the TCF of discs. There is only one clear peak, corresponding to scales close to the semi-minor axis: because it probes equilateral triangles, the function seems to pick up elongated ellipses as rows of discs of radius $\beta$. The possibility to discern two peaks in the TCF in the presence of different bubble sizes will be limited by two factors. First, the number of bubbles, as we have seen that too many or too few objects leads to a weak signal. We find that when the ratio of large over small bubble numbers exceeds 10 or drops below 0.1, one of the two peaks flattens out. Second, the separation between the two radii: the TCF peaks of bubbles with $R=6~\mathrm{px}$ and $R=10~\mathrm{px}$ will overlap into one single wide peak and we will not be able to differentiate them anymore.

Analytic derivations predict a log-normal distribution for the bubble sizes during reionisation, which becomes increasingly peaked as bubble grow and merge \citep{furlanetto_2004_gaussianity}. This first result has been confirmed by many authors looking at simulations  \citep{mcquinn_2007_HIIregions,zahn_2007_SPA,lin_2016_review}. Let two boxes where the bubble centres are randomly distributed, but the radius sizes are sampled from a log-normal distribution. Because we generate the box, we know the exact distribution of bubble radii within and we want the TCF to trace it. On Fig. \ref{fig:radius_distri}, we compare the TCF (solid line) with the actual radius distribution (histogram) for 2 different bubble size distributions: on centred on $R = 5~\mathrm{px}$ with variance $2~\mathrm{px}$, corresponding to the early stages of EoR, and one with $R=11~\mathrm{px}$ with variance $1~\mathrm{px}$, corresponding to later stages. The number of bubbles is adjusted to reach a filling fraction $x_\ion{H}{II}=0.09$. Overall, there is a reasonably good match between the two. We see that as the mean radius increases, the discrepancy between peaking scale and real bubble size grows in a way that the linear relation of Eq. \ref{eq:rvsR_binary} cannot fully compensate for. Note that results are similar for a Gaussian distribution of the bubble sizes.

\subsection{Results on 21CMFAST simulation}
\label{sec:21cmFAST}

\begin{figure*}
    \centering
    \includegraphics[height=0.185\textheight]{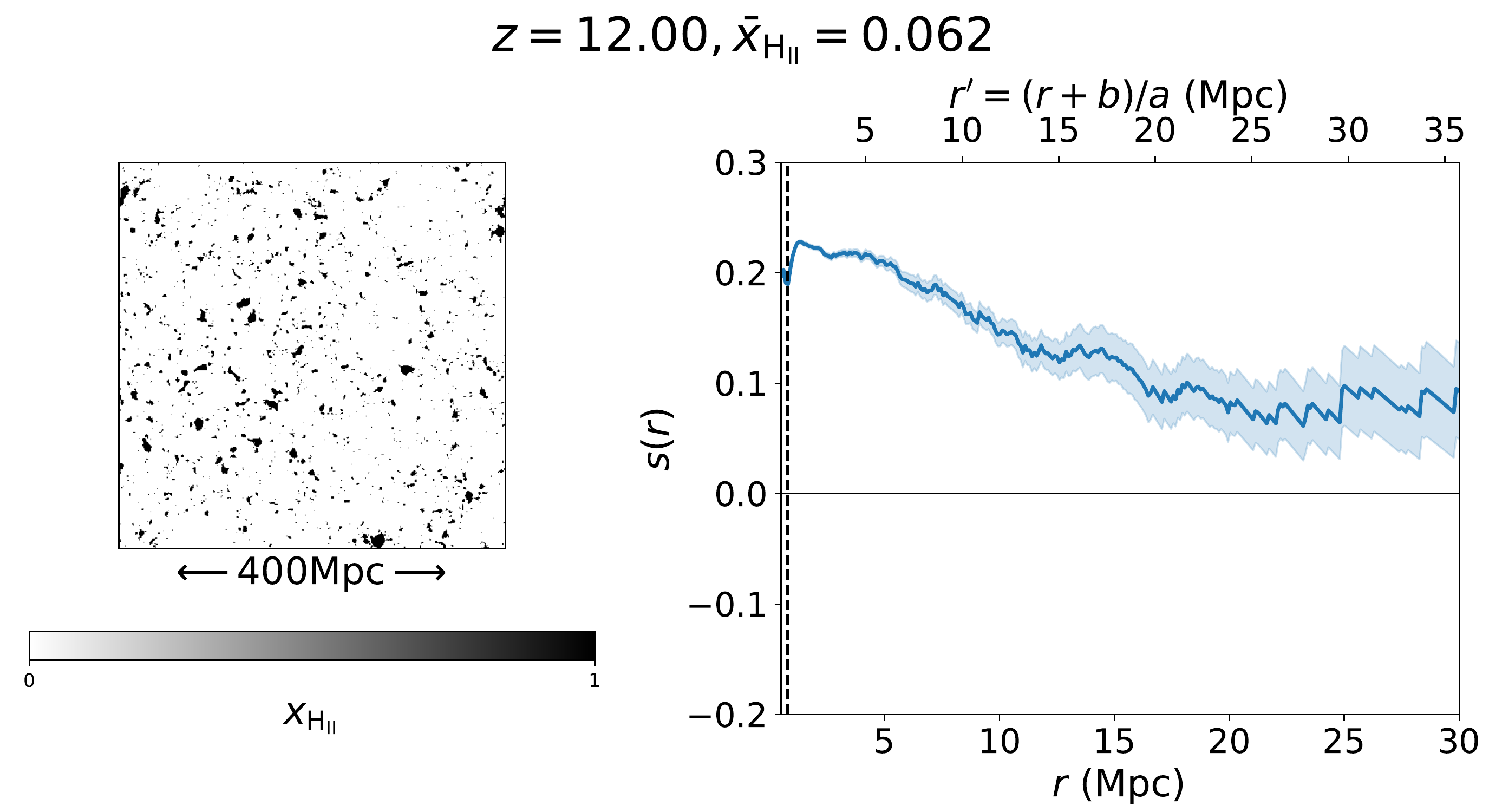}
    \includegraphics[height=0.185\textheight]{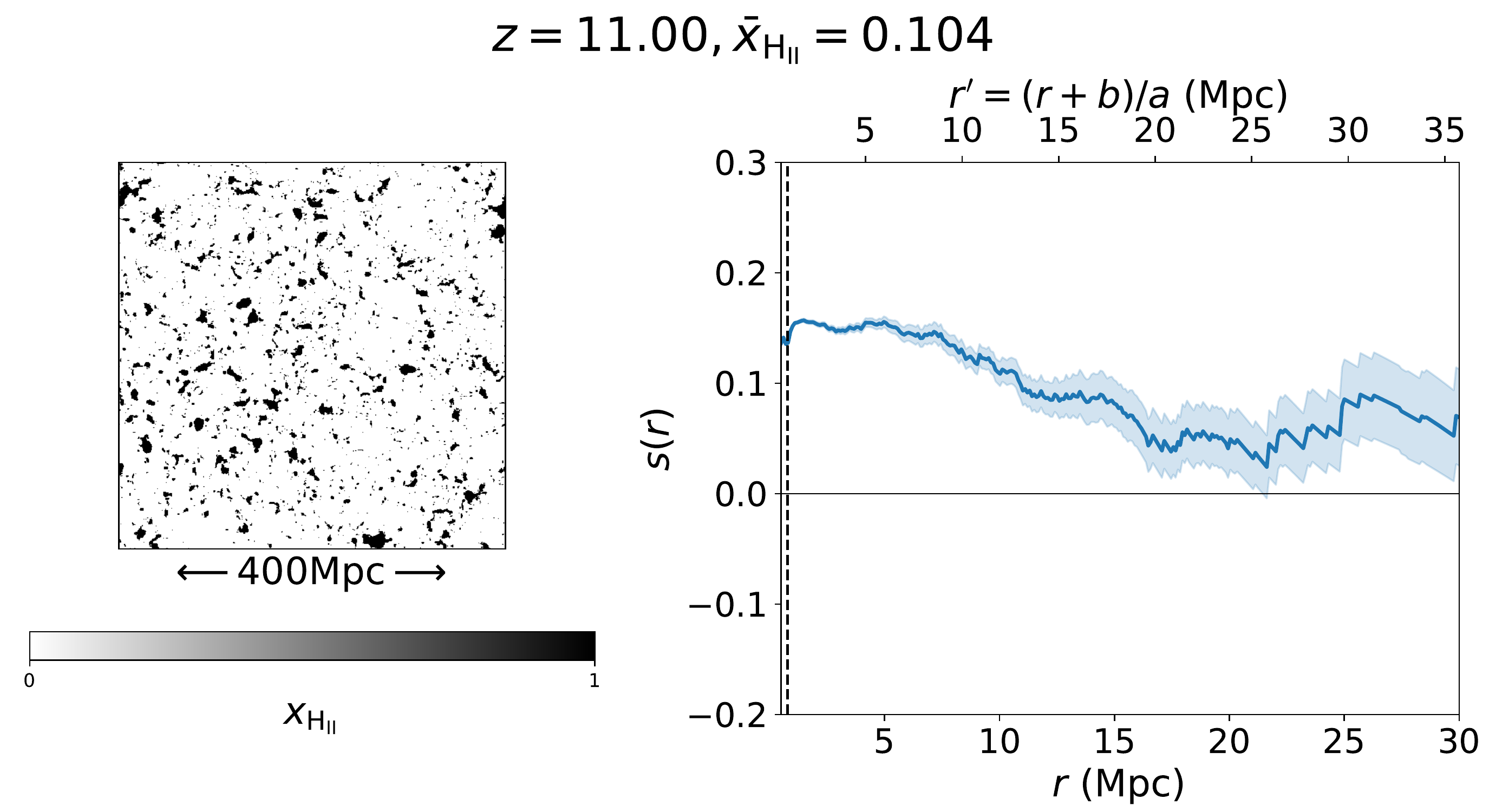}\\
    \includegraphics[height=0.185\textheight]{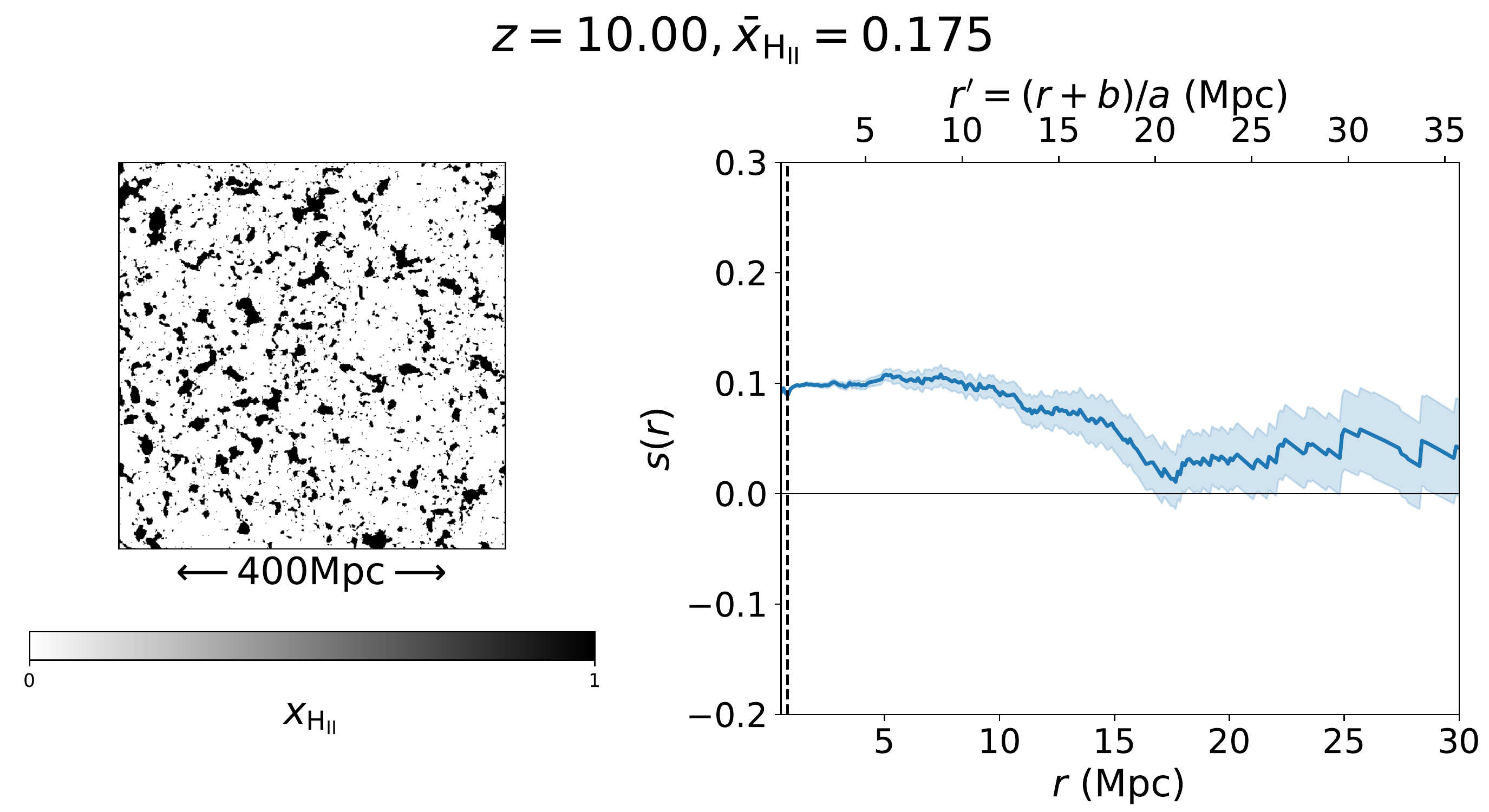}
    \includegraphics[height=0.185\textheight]{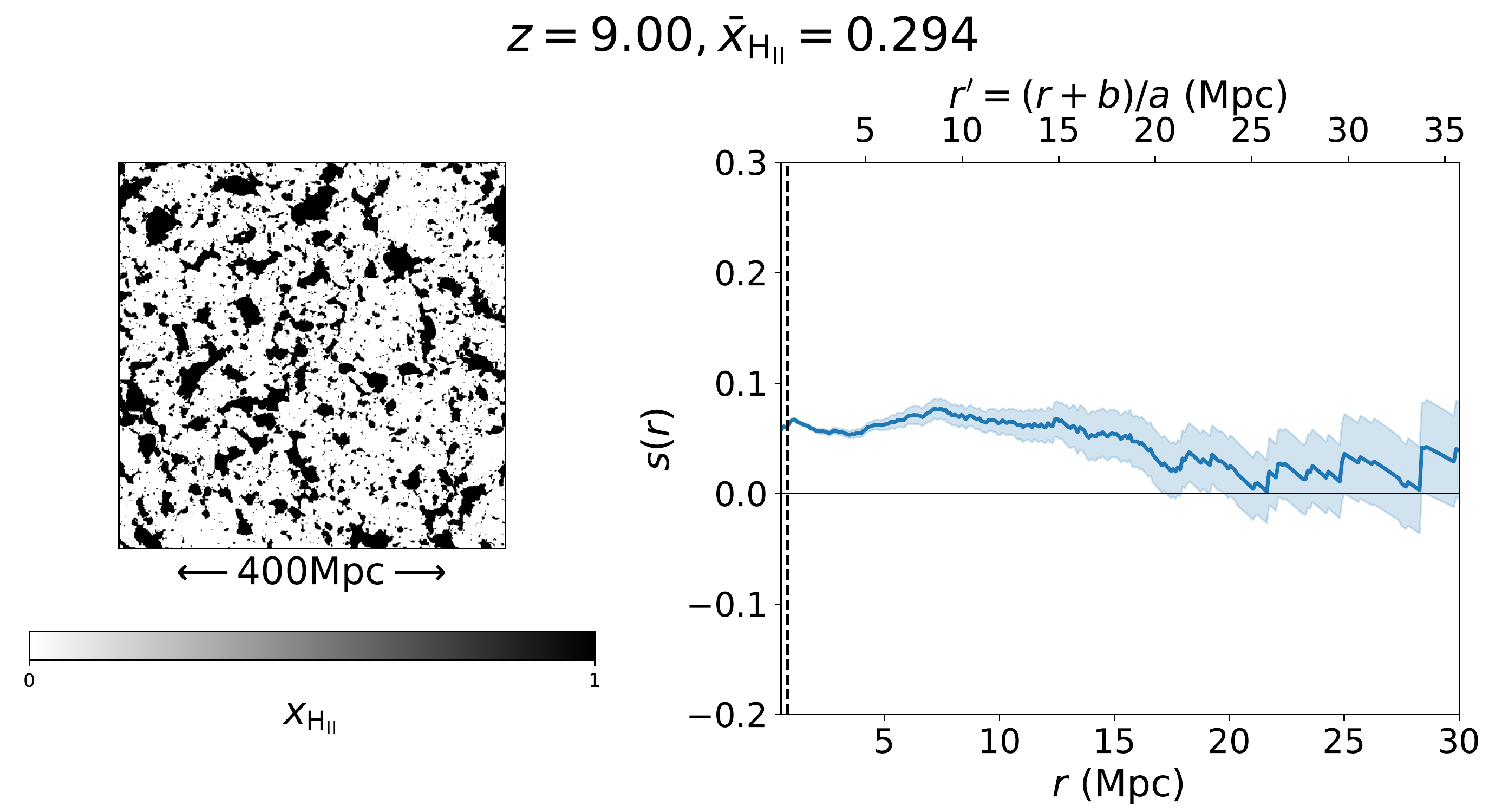}\\
    \includegraphics[height=0.185\textheight]{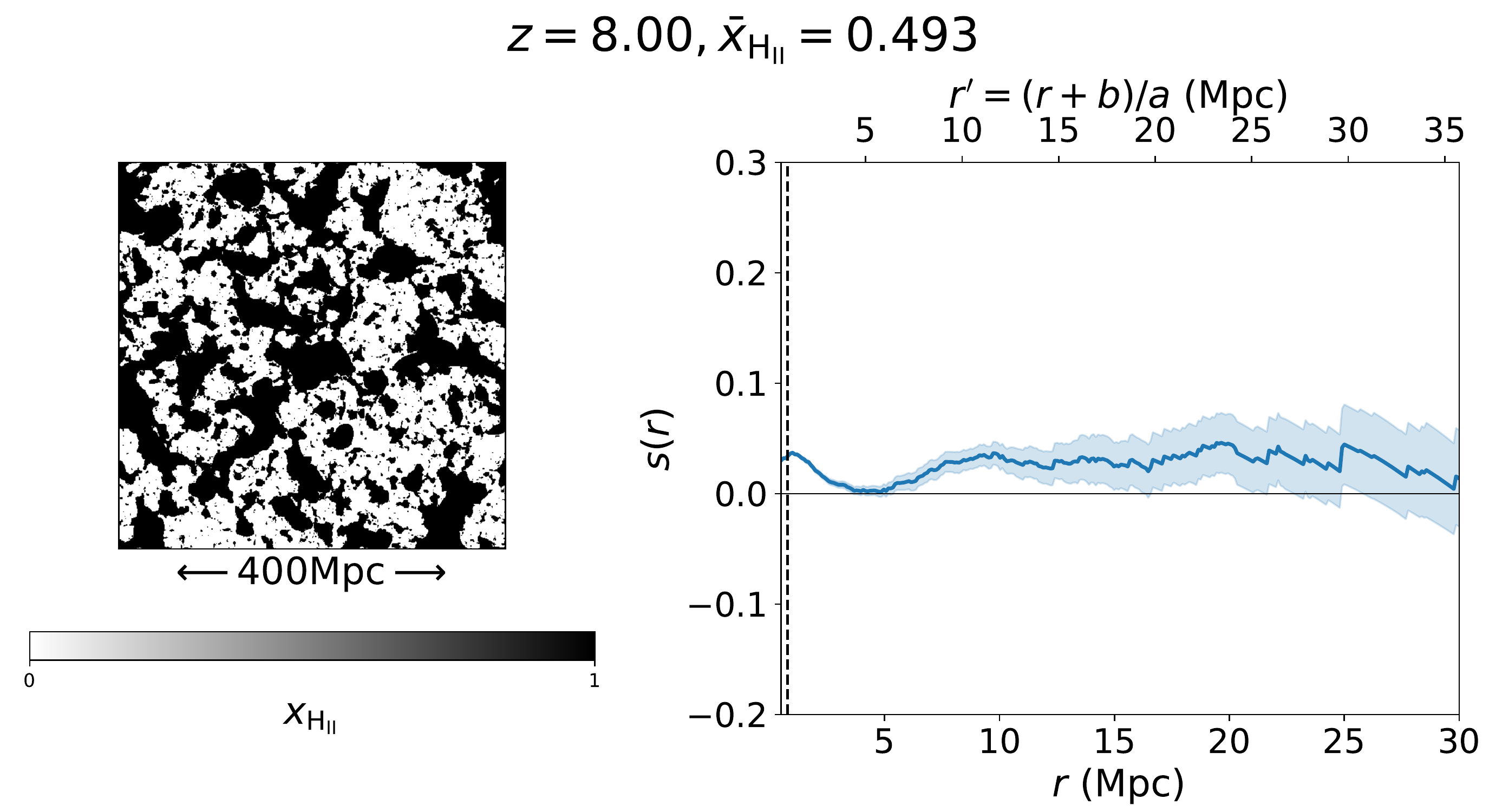}
    \includegraphics[height=0.185\textheight]{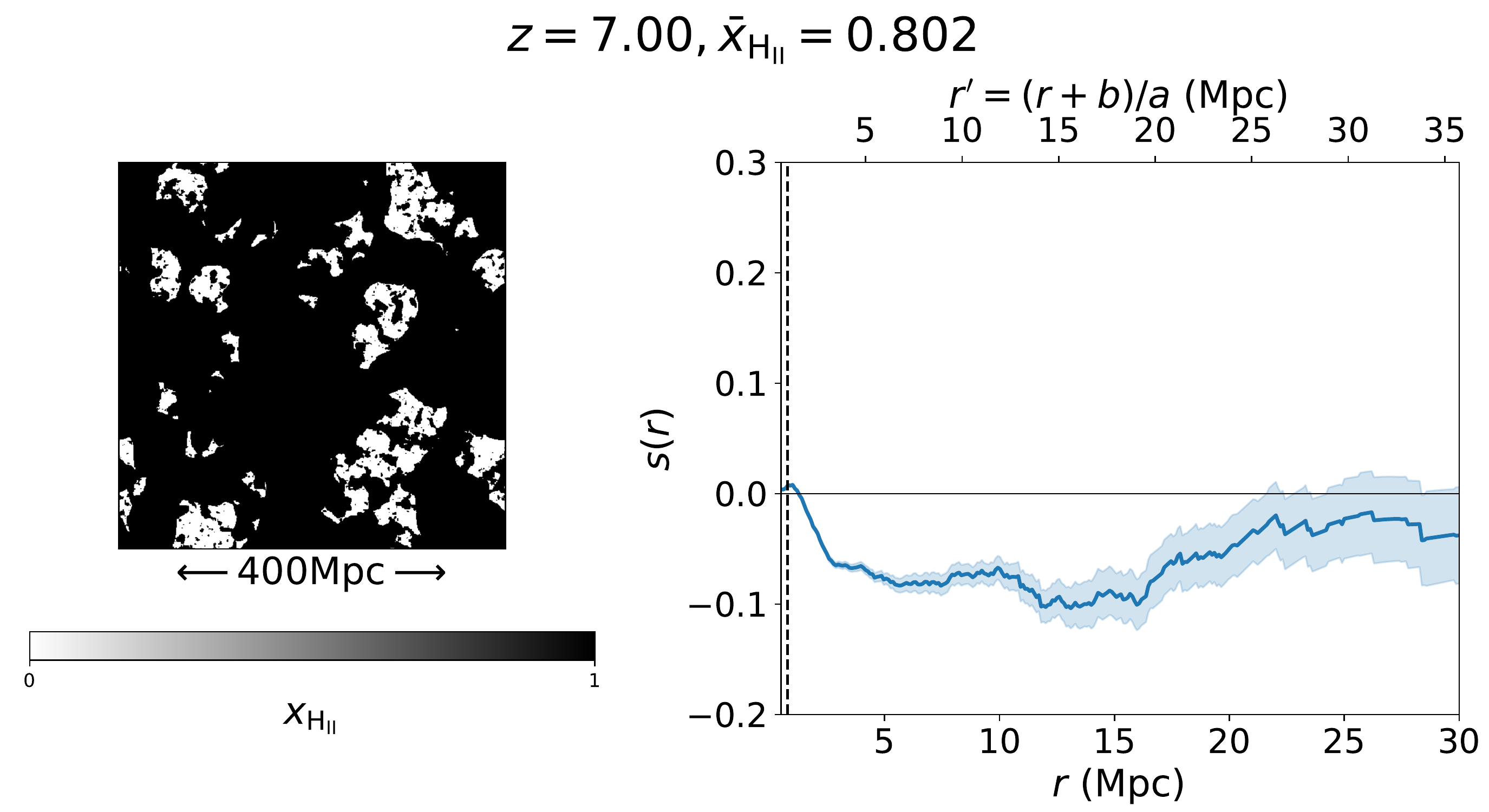}\\
   \includegraphics[height=0.185\textheight]{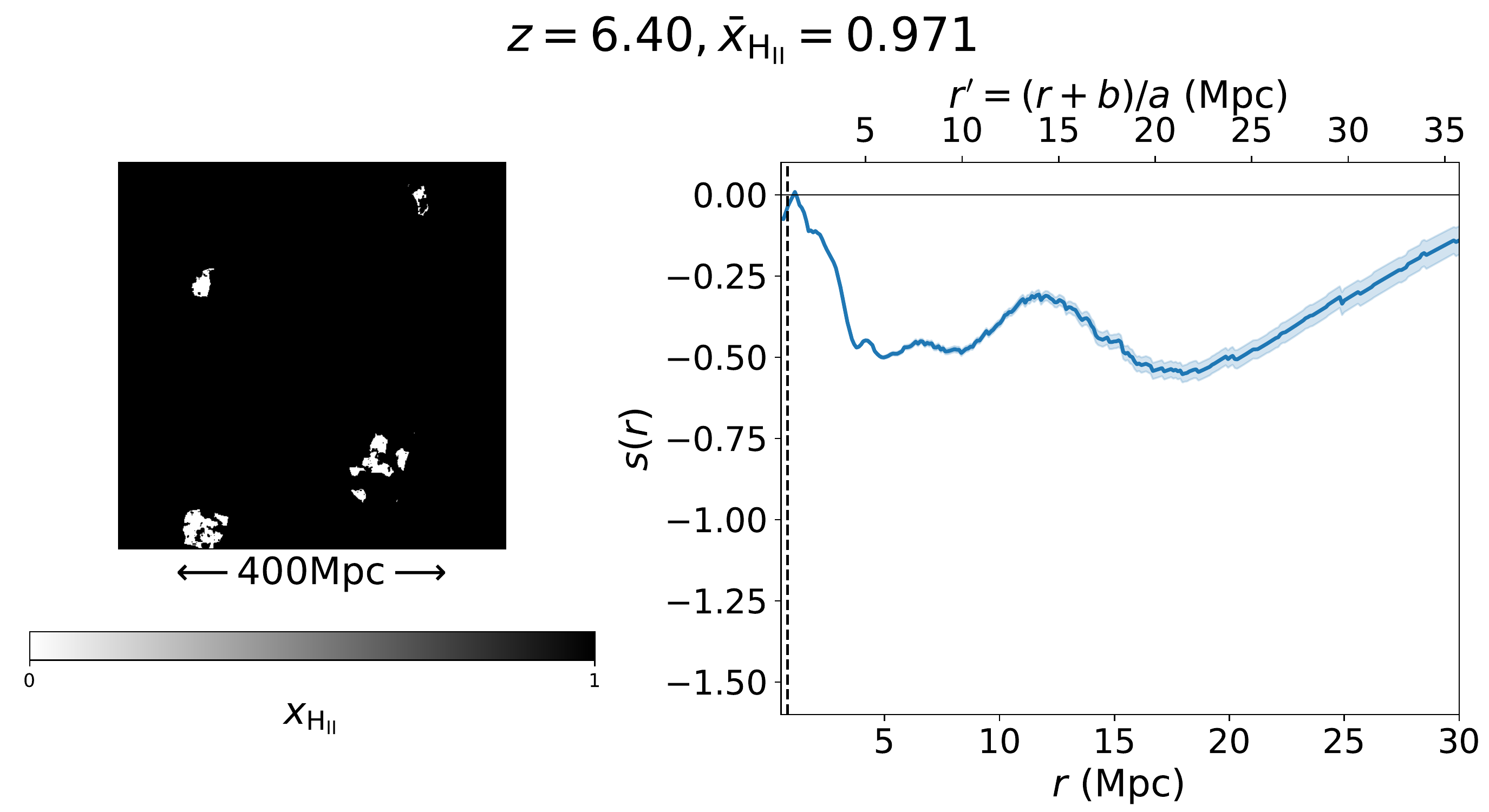}
    \includegraphics[height=0.185\textheight]{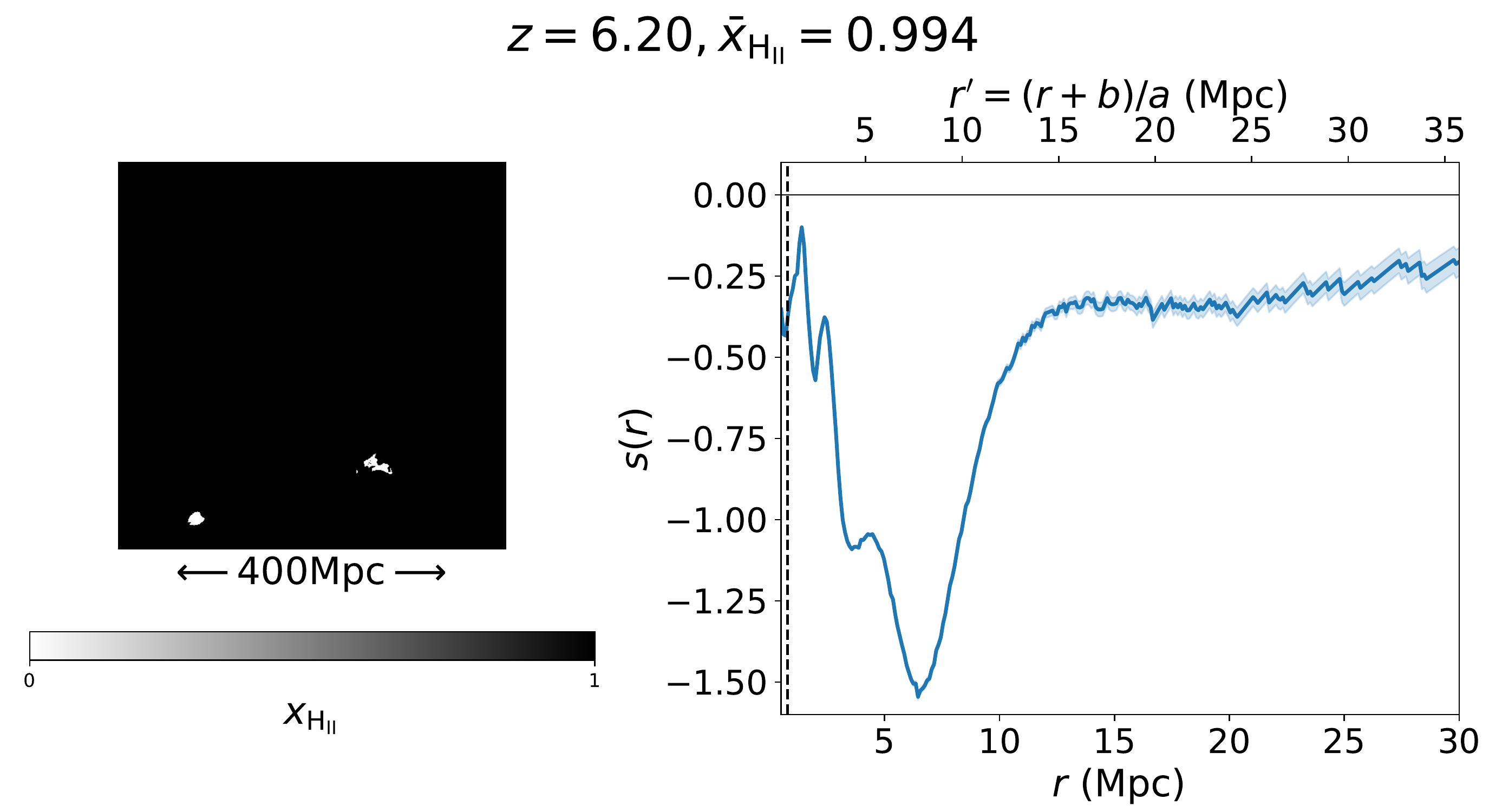}
    \caption{Triangle correlations for 2D slices of our simulation at various redshifts. The upper $x$-axis is the lower one scaled by the linear relation in Eq. \ref{eq:rvsR_binary}. Error bars correspond to the variance estimated from a Gaussian random field of same dimensions.}
    \label{fig:21cmFAST_SC_variousz}
    \includegraphics[height=0.18\textheight]{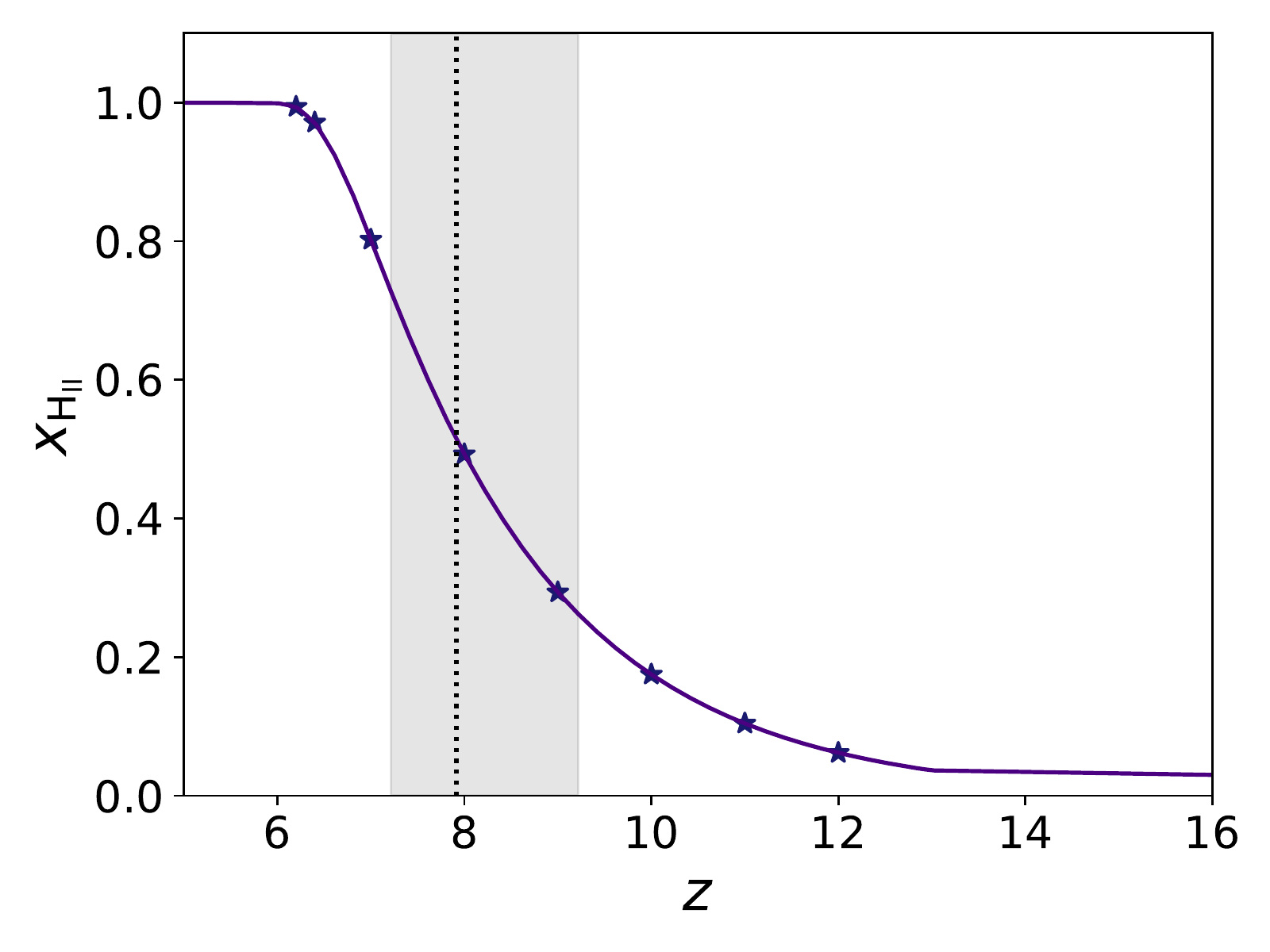} 
    \caption{Reionisation history of our simulation. The dotted line corresponds to the midpoint of reionisation, i.e. $z=7.9$, and the shaded region to global ionised fraction between $25\%$ and $75\%$. The starred points are the stages of reionisation represented above.}
    \label{fig:21cmFAST_reio_history}
\end{figure*}

\begin{figure*}
\centering
\includegraphics[width=0.24\textwidth]{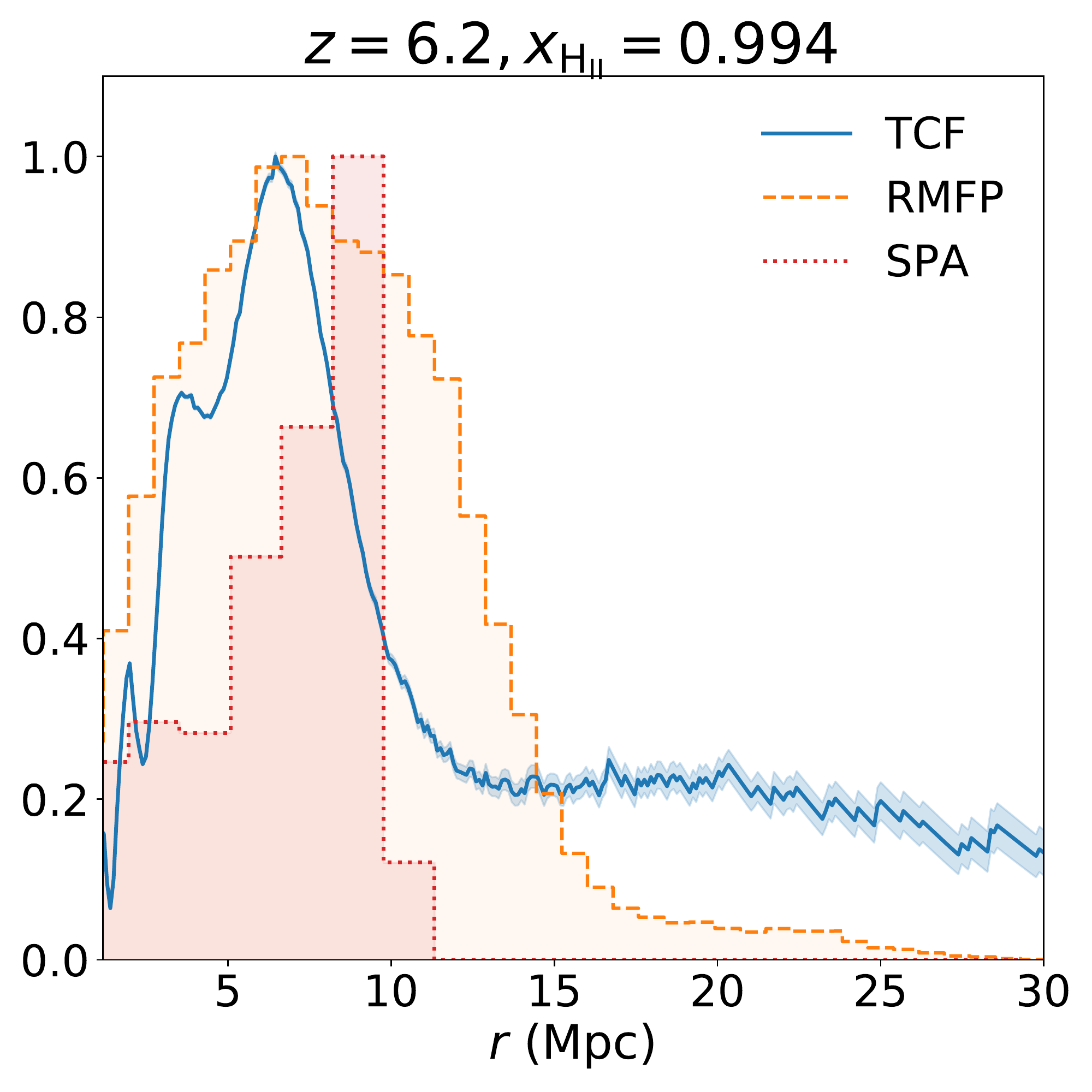}
\includegraphics[width=0.24\textwidth]{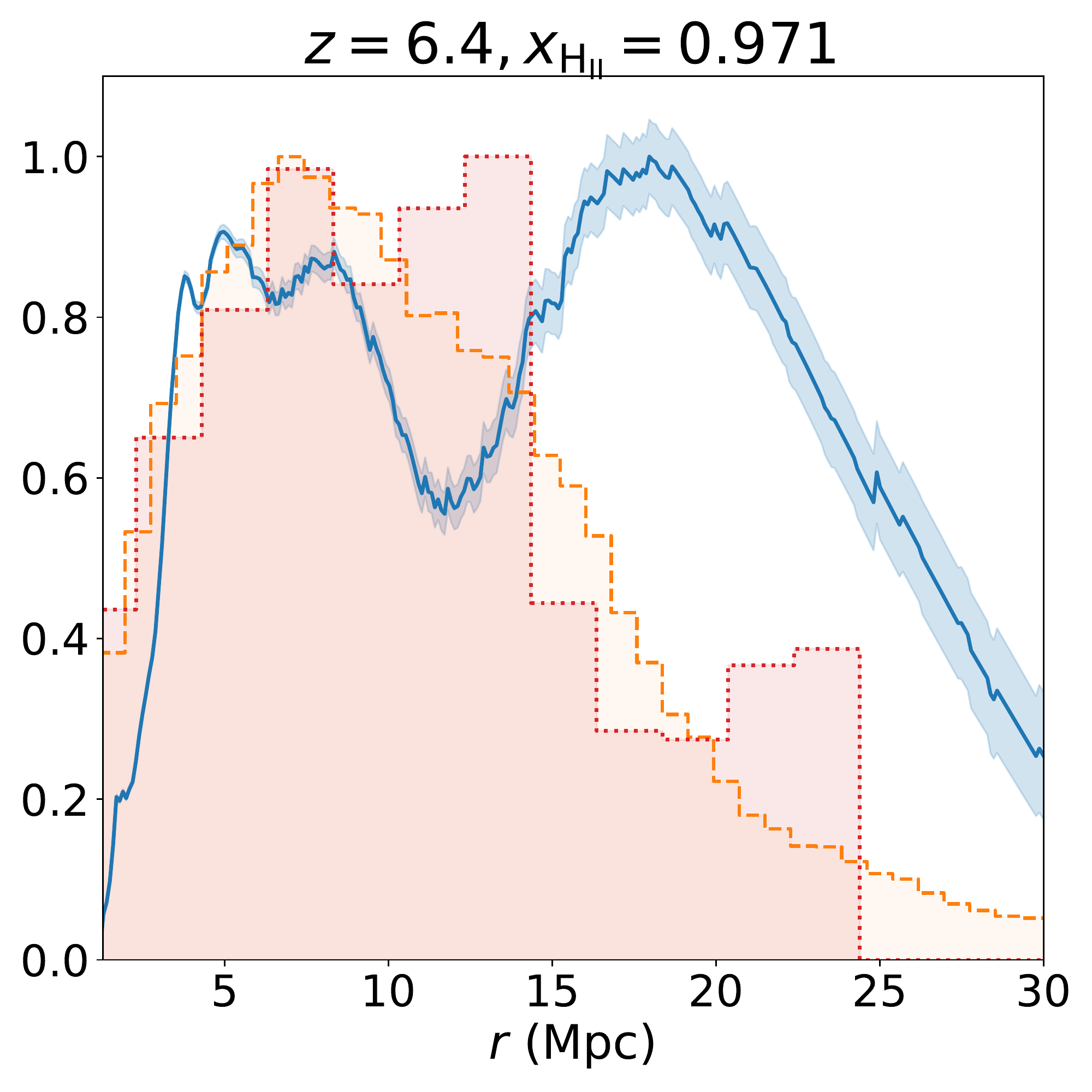}
\includegraphics[width=0.24\textwidth]{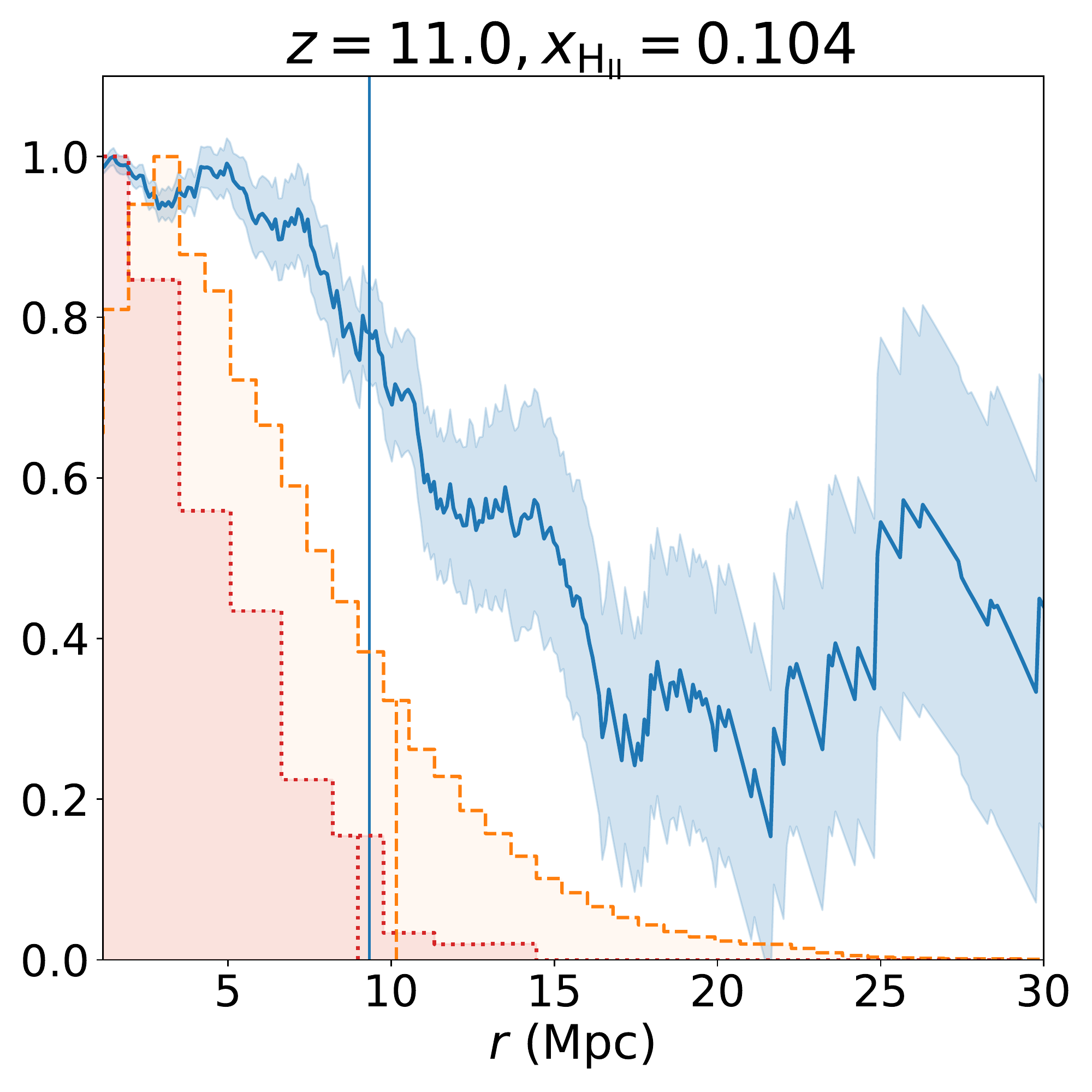}
\includegraphics[width=0.24\textwidth]{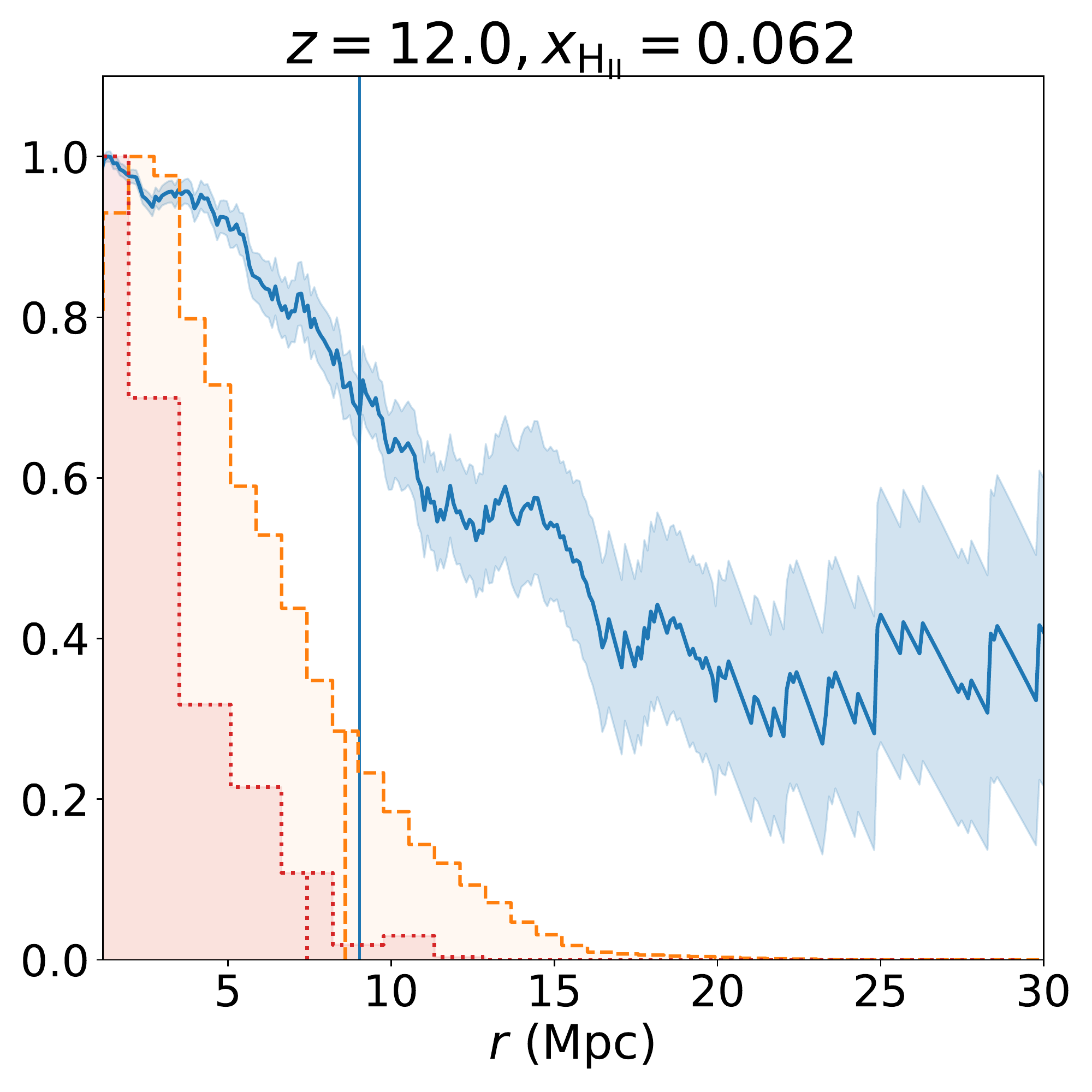}
\caption{Comparison of different methods to estimate the size of neutral (for $z=6.2$ and 6.4) or ionised ($z \geq 11$) regions in our 21CMFAST simulation. Error bars correspond to the variance estimated from a Gaussian random field of same dimensions.}
\label{fig:21cmFAST_method_comparison}
\end{figure*}

Now that our method has been tested and characterised on toy models, we apply it to more physical ionisation fields, extracted from the semi-numerical simulation of reionisation 21CMFAST\footnote{\url{http://github.com/andreimesinger/21cmFAST}} \citep{21cmFAST_2007,21cmFAST_2011}. We choose 21CMFAST because, thanks to relatively short executing times, it gives a large flexibility in parametrisation. It also includes an implementation of the random mean free path algorithm to estimate the bubble size distribution of data cubes, which we can compare to our estimator. Note that an important difference with the toy models is that the bubble locations are no longer random, and the TCF might pick up correlations between the bubble centres on large scales. The 21CMFAST code first generates a high-redshift linear density field which is then evolved to lower redshifts using linear theory and the Zel'dovich approximation. The ionisation field is extracted from this density field using excursion-set theory for haloes with virial temperature $T > 10^4~\mathrm{K}$. We use this code to generate a box with sufficient resolution to obtain a $\ion{H}{I}$ field of $512^3$ pixels and side length $L=400~\mathrm{Mpc}$ for the following cosmology: $\Omega_m=0.308$, $\Omega_b=0.049$ and $H_0 = 67.74~\mathrm{km.s}^{-1} \mathrm{Mpc}^{-1}$ \citep{planck_2015_cosmo_params}. The resulting reionisation history has its midpoint at $z=7.9$ for a duration of $\Delta z = z \left(x_\ion{H}{II} = 0.10 \right) - z \left( x_\ion{H}{II}=0.99 \right) = 4.9 $ and gives an integrated Thomson optical depth of $\tau = 0.067$ (see Fig. \ref{fig:21cmFAST_reio_history}).

The output of the simulation is a 3D $\ion{H}{I}$ field but we choose to analyse 2D slices to be closer to actual observations. We also convert the given $\ion{H}{I}$ field into a binary $\ion{H}{II}$ field in order to get positive triangle correlations when most of the field is ionised and negative correlations when the sky is mainly neutral (see Eq. \ref{eq:def_bspectrum} for $x_\ion{H}{I} = 1 - x_\ion{H}{II}$ and discussion at the end of Sec. \ref{sec:mathematical_formalism}), in continuity with previous sections. Fig. \ref{fig:21cmFAST_SC_variousz} presents the TCF of 2D slices at redshifts $z=12.0, 11.0, 10.0, 9.0, 8.0, 7.0, 6.4$ and $6.2$ corresponding to global ionisation levels of $x_\ion{H}{II} = 0.09, 0.10, 0.18, 0.29, 0.49, 0.80, 0.97$ and $0.99$ respectively, along with a picture of the corresponding $\ion{H}{II}$ field. For each simulation, results are shown with variance estimated from the TCF of 20 Gaussian random fields (GRF) of same dimensions. This indicates that the signal is more significant at small scales, a point already mentioned in Section \ref{subsec:scales_random} and gives an idea of how well this field differs from a GRF. However, these errors will not take into account the non-Gaussian nature of the field if the 2D slice considered is not representative of the overall field, especially at high redshifts where the signal is highly non Gaussian. Indeed, consider the $z=6.2$ slice on Fig. \ref{fig:21cmFAST_SC_variousz}: the small ionised regions seen in the field are isolated features. If we compute triangle correlations for more realisations of this simulation, it is unlikely that the exact same feature will appear in the field and so in the TCF. Therefore there will be more variance on small scales than the error bars on the Figure. Ideally, one would find a theoretical expression for the covariance of the TCF, but we keep this derivation for future work.

At high redshift, in what is often referred to as the \textit{pre-overlap phase}, we see on the TCFs of Fig. \ref{fig:21cmFAST_SC_variousz} that there is more power at small scales but no clear peak. This is likely due to the variety of $\ion{H}{II}$ regions: many small ionised regions around young sources parcel the neutral background out. However, by looking at the signal-to-noise ratio (SNR), we can infer an upper limit on the sizes of the ionised regions: at $z=13$, scales smaller than $8.9~\mathrm{Mpc}$ (after conversion with Eq. \ref{eq:rvsR_binary}) contribute for $80\%$ of the cumulative SNR; at $z=12$ and $z=11$, this upper limit increases respectively to $10.8~\mathrm{Mpc}$ and $11.2~\mathrm{Mpc}$. The structure of the ionisation field at high redshift directly relates to the way the 21CMFAST algorithm is constructed and particularly to the use of excursion set theory. In excursion-set theory, when the average density of a region exceeds a given threshold, it collapses to form a halo. When applied to reionisation, the threshold additionally considers ionising photons production: if the region has produced a sufficient number of ionising photons with respect to its mass and volume, then it is considered ionised. Because of this, the ionisation field in the early stages of reionisation will have many very small sources rather than a few efficient sources. It would be interesting to compute the TCF of other reionisation simulations, in particular simulations with more efficient escape fractions \citep{seiler_2019}. This also includes simulations where reionisation is led by Active Galactic Nuclei (AGN): ionising sources are more scarce and have a better ionising efficiency, leading to a topology more similar to the toy models used in Sec. \ref{subsec:scales_random}. In this perspective, our method may be able to differentiate between different reionisation scenarios. Such a study goes beyond the scope of this work but we refer the interested reader to \citet{bispectrum_xray_2019}, where the authors use the bispectrum as a probe for non-Gaussianities due to X-ray heating.

Later on, when the global ionisation fraction reaches values between $25\%$ and $75\%$, negative signal coming from neutral regions overlaps with the positive signal from ionised regions. The TCF flattens, and therefore cannot give information about the morphology of the field. However, measuring a flat signal from actual data could be interpreted as the reionisation process being in its middle stages.
For $z>7$, most of the sky is ionised and only a few remote neutral islands remain. This is the \textit{post-overlap phase}. We see on Fig. \ref{fig:21cmFAST_SC_variousz} that the sizes of these neutral islands are efficiently picked up by the TCF. For $z=6.2$, there is a very clear negative peak at scales $r = 7.7~\mathrm{Mpc}$ which correspond to a radius size of $r' = 9.2~\mathrm{Mpc}$ once the linear relation in Eq. \ref{eq:rvsR_binary} is applied as a correction. If we roughly estimate the size of the neutral zones, we find that they are about $\sim 11~\mathrm{px} \sim 8.6~\mathrm{Mpc}$ in radius, which is very close to the estimation given by triangle correlations. For $z=6.4$, we see two clear peaks in the signal, corresponding to the two sizes of ionised regions seen in the corresponding real space field. The first peak is spread over scales $4.6 < r < 7.8~\mathrm{Mpc}$ i.e. $5.5 < r' < 9.3~\mathrm{Mpc}$, while the second one is more narrow, centred around $20.2~\mathrm{Mpc}$. For comparison, we find that we can sieve the two larger neutral islands in the field with disks of radius $\sim 23~\mathrm{px} \sim 18~\mathrm{Mpc}$; whereas the smaller ones can fit in disks whose radii range from $7~\mathrm{Mpc}$ to $12~\mathrm{Mpc}$. This motivates a further study of how well the TCF performs compared to common BSD algorithms. 

\begin{table}
	\centering
	\caption{Peaking scale (in Mpc) for different methods to obtain the bubble size distribution of realisations of the 21CMFAST simulation at different redshifts.}
	\label{tab:3_methods}
	\begin{tabular}{c|c|c|c|c|} % four columns, alignment for each
		 $z$ & $x_\ion{H}{II}$ & TCF & RMFP & SPA \\
		\hline
		13.0 & 0.036 & < 8.9 & 2.01 & < 5.9 \\
		12.0 & 0.062 & < 10.8 & 2.34 & < 7.4 \\
		11.0 & 0.010 & < 11.2 & 3.12 & < 8.9 \\
		6.4  & 0.971 & 6.2 \& 20.2 & 7.0 & 13.9\\
		6.2  & 0.994 & 7.7 & 7.0 & 9.4 \\
		\hline
	\end{tabular}
\end{table}

Let's consider the spherical average (SPA) and the random mean free path (RMFP) methods. Note that these two algorithms need to be applied to real space images whereas the TCF is used on Fourier data. We choose not to compare our results to the friend-of-friends algorithm \citep{iliev_2006} because it considers overlapping bubbles as a unique large ionised region, which is fundamentally different from our approach. We use versions of SPA and RMFP implemented by the authors of \citet{giri_2018_bubble_sizes}\footnote{\texttt{tools21cm} are found on \url{https://github.com/sambit-giri/tools21cm}.}. Recall that the SPA algorithm looks for the largest sphere around each ionised region whose ionisation level exceeds a given threshold, here chosen to be $x_\ion{H}{II}=0.5$. RMFP, on the other side, relies on MCMC: it looks for the first neutral cell encountered in a random direction starting from an ionised pixel and records the length of the ray to build a histogram. The SPA and RMFP outputs are probability distributions of radii.%, not a delta function at the exact scale: they are called \textit{diffusive} estimators of the bubble size. %The RMFP histogram will be sensitive to the choice in the threshold chosen to distinguish between ionised and neutral pixels \citep{friedrich_2011_topology_tomography}; but this is not an issue here since the ionisation field we use is already binary.  
Results can be seen on Fig. \ref{fig:21cmFAST_method_comparison} for $z=6.2, 6.4, 11.0$ and 12.0 and are detailed in Table \ref{tab:3_methods}. For clarity, we plot $-s(r)$ in the first two panels to have a positive signal for all redshifts. At low redshift, our method mostly agree with the other BSDs, although the TCF exhibits a narrower peak; we are also able to clearly distinguish between two characteristic scales for $z=6.4$, whereas SPA et RMFP only give one most likely scale, apparently corresponding to the smaller neutral regions. At higher redshift, the SPA distributions tend to $r \sim 0~\mathrm{Mpc}$ and we choose to compare values of quantiles rather than maximum likelihood values: at $z=11$, according to SPA, there is a $80\%$ probability that the ionised regions have a radius smaller than $13.3~\mathrm{Mpc}$; similarly, scales $< 11.2~\mathrm{Mpc}$ represent $80\%$ of the cumulative SNR for the TCF. These upper limits, computed for both $z=11$ and $z=12$, are shown as vertical lines on Fig. \ref{fig:21cmFAST_method_comparison}. RMFP gives a maximum likelihood radius of $3.12 \pm 0.78~\mathrm{Mpc}$ (the error corresponding to the bin size), therefore the most quantitative result. For all three methods, the upper limit increases as redshift decreases, which corresponds to ionised regions growing with time. Note that the upper limit given by SPA is the lowest: our results corroborate those of \citet{lin_2016_review}, who argue that SPA is heavily biased towards smaller bubble radii than the actual bubble sizes. Indeed, they find that for a single bubble of radius $R$, the SPA probability distribution peaks at $R/3$. On the contrary, they find the RMFP method to be unbiased and to peak at the correct bubble size. Note that on the two right panels of Fig. \ref{fig:21cmFAST_method_comparison}, we see some signal at large scales. This can be related to statistical noise on one side -- we analyse a unique slice of the simulation; and to the potential correlations between separate bubbles, which are not randomly distributed anymore.

Overall, our estimator performs well with respect to comparable BSD algorithms. However, for this method to be an useful tool in the analysis of upcoming radio observations, a deeper analysis, including the study of different types of reionisation simulations, would be required. In practise, we would also use this method as a forward modelling process, and compare triangle correlations from observations to a set of predicted signals for different types of ionisation fields. We keep the construction of such a database for future work.

\section{Relation to observations}
\label{sec:relation_obs}

\subsection{Visibilities and closure phases}
\label{subsec:interferometry}

When gathering data with an interferometer, the signal will be measured for pairs of antennae separated by a baseline $D$. The measurement is made in terms of a complex visibility $V(u,v)$, where $u$ and $v$ are the projection of the baseline in wavelength units on the plane perpendicular to the vector $\bm{s}_0$ pointing to the phase reference position, i.e. the centre of the field to be imaged. %There should be a third coordinate, along $\bm{s}_0$, in the definition of the visibility, but we will see that it can be ignored most of the time.
In Eq. \ref{eq:brithness21cm}, the brightness temperature depends on the position in the sky $\bm{r}$ where the signal is observed: $\delta T_\mathrm{b} (\bm{r},z) \propto \, x_\ion{H}{I}\, (\bm{r},z) $. $\delta T_\mathrm{b}$ is the intensity $I_\nu$ of the 21cm signal observed through the interferometer for a given frequency $\nu$ (i.e. for a given redshift, since $\nu = \nu_{21\mathrm{cm}}/(1+z)$)\footnote{We assume the instrument to have narrow bandpass filters and so to probe exactly $\nu$ rather than a frequency band centred on $\nu$.}. In what follows, we will use $I_\nu (\bm{r}) = \bar{I}_\nu \, x_\ion{H}{I} (\bm{r},z)$ and $\delta T_\mathrm{b}(\bm{r},z)$ interchangebly. If the interferometer probes a sufficiently small region of the sky compared to the beam width of the antennae, we can approximate this region by a flat plane. Then the source intensity distribution is a function of two real spatial variables $(l,m)$ and the van-Cittert theorem tells us that the complex visibility is the 2D inverse Fourier transform of the intensity, corrected by the normalised average effective collecting area $A(l,m)$ of the two antennae \citep{book_interferometry_thomson}:
\begin{equation}
    A(l,m)\, I_\nu(l,m) = \sqrt{1-l^2-m^2} \iint V(u,v)\, \mathrm{e}^{2i\pi \, (ul+vm)} \ \mathrm{d}u \, \mathrm{d}v,
\end{equation}
that we can re-write as
\begin{equation}
    V(u,v) = \Hat{A}^\star(u,v) * \Hat{I}_\nu (u,v)
\end{equation}
where $*$ denotes a convolution, $\Hat{A}$ and $\Hat{I}_\nu$ are respectively the 2D Fourier transforms of $A$ and $I_\nu$ and we define $A^\star = A/\sqrt{1-l^2-m^2}$.
Note that this approximation introduces a phase error that may need to be taken into account when applying our method, based on phase information. Because $A(l,m)$ is a measurable instrumental characteristic, if we know our instrument sufficiently well, we can deconvolve the measured complex visibilities by $\Hat{A}^\star$ and obtain, after correcting for the different pre-factors, the $\Hat{x}(\bm{k})$ terms to compute the triangle correlation function. 

One of the interests methods based on phase information is that the phases of the measured visibilities, combined in a bispectrum, will not be sensitive to errors related to calibration \citep{jennison_1958_fourier_phases,monnier_phases_interferometry}. Consider the visibility $V_{ij}^\mathrm{m}$ measured between two antennae $i$ and $j$ at a given frequency $\nu$. It will have contributions from the true visibility $V_{ij}^{\mathrm{true}}$, coming from the cosmological signal, but also from the amplitude and phase errors of each antenna, modelled by a complex gain $G_{i} = \vert G_{i} \vert \, \mathrm{e}^{i\phi_{i}}$, such that:
\begin{equation}
\begin{aligned}
    V_{ij}^{\mathrm{m}} & = G_{i} G_j^* \ V_{ij}^{\mathrm{true}},\\
    & = \vert G_i G_j \vert \ \mathrm{e}^{i(\phi_i - \phi_j)} \ V_{ij}^{\mathrm{true}},
\end{aligned}
\end{equation}
where $*$ denotes the complex conjugate.
The amplitude of this gain will come from beam specific effects such as mirror reflectivity, detector sensitivity or local scintillations whereas the phase term can originate either from telescope errors or from outside effects such as atmospheric turbulence \citep{levrier_2006,monnier_phases_interferometry}. If we combine the signal from three antennae forming a closed triangle, we can avoid this phase error and we will be left only with what is called the closure phase. Indeed, consider three baselines $ij$, $jk$ and $ki$ observing at the same given frequency $\nu$. We can write the bispectrum of their complex visibilities as
\begin{equation}
\label{eq:closure_relation}
\begin{aligned}
    \mathcal{B}_{ijk} & = V_{ij}^\mathrm{m} \, V_{jk}^\mathrm{m} \, V_{ki}^\mathrm{m} \\
    & = \vert G_i G_j  G_k \vert^2 \ \mathrm{e}^{i(\phi_i - \phi_j)} \, \mathrm{e}^{i(\phi_j - \phi_k)} \, \mathrm{e}^{i(\phi_k - \phi_i)} \ V_{ij}^\mathrm{true} \, V_{jk}^\mathrm{true} \, V_{ki}^\mathrm{true} \\
     & = \vert G_i G_j G_k \vert^2 \ V_{ij}^\mathrm{true} \, V_{jk}^\mathrm{true} \, V_{ki}^\mathrm{true},
\end{aligned}
\end{equation}
where in the last step, the different phase terms cancel each other out: the phase of the measured bispectrum is the phase of the true bispectrum. By construction, the bispectrum $\mathcal{B}(\bm{k},\bm{q})$ in Eq. \ref{eq:def_bspectrum} considers a closed triangle configuration, for three vectors $\bm{k}$, $\bm{q}$ and $-\bm{k} - \bm{q}$. Because we choose to work in two dimensions, the three vectors lie in the same plane on the sky, perpendicular to the line-of-sight, and are measured for the same frequency so that the closure relation holds. We will then be able to use our method on observational data without worrying about calibration errors.  For an example of the use of bispectrum closure phases in interferometry, we refer the reader to \citet{thyagarajan_2018_bispectrum_phases}. In this work, the authors compare the bispectrum phase spectra, i.e. the phases of the bispectrum $\mathcal{B}_{ijk}$ in Eq. \ref{eq:closure_relation}, coming from different components of a simulated signal, i.e. a single point source, diffuse foregrounds and $\ion{H}{I}$ fluctuations from the EoR and demonstrate that a quantitative relationship exists between the EoR signal strength and the whole bispectrum phase power spectra.

These properties are a major benefit of our technique compared to other solutions found in the literature to estimate the characteristic size of ionised regions from interferometric data as these always require to reconstruct the real-space image corresponding to observations. Indeed, our method can directly use as input the complex visibilities observed by an interferometer, and if we choose to use closure phases, results will be independent of antenna-based calibration and calibration errors.  However, in practise, there are some limitations to the use of the closure relation. 

\begin{figure}
\centering
\includegraphics[width=.3\textwidth]{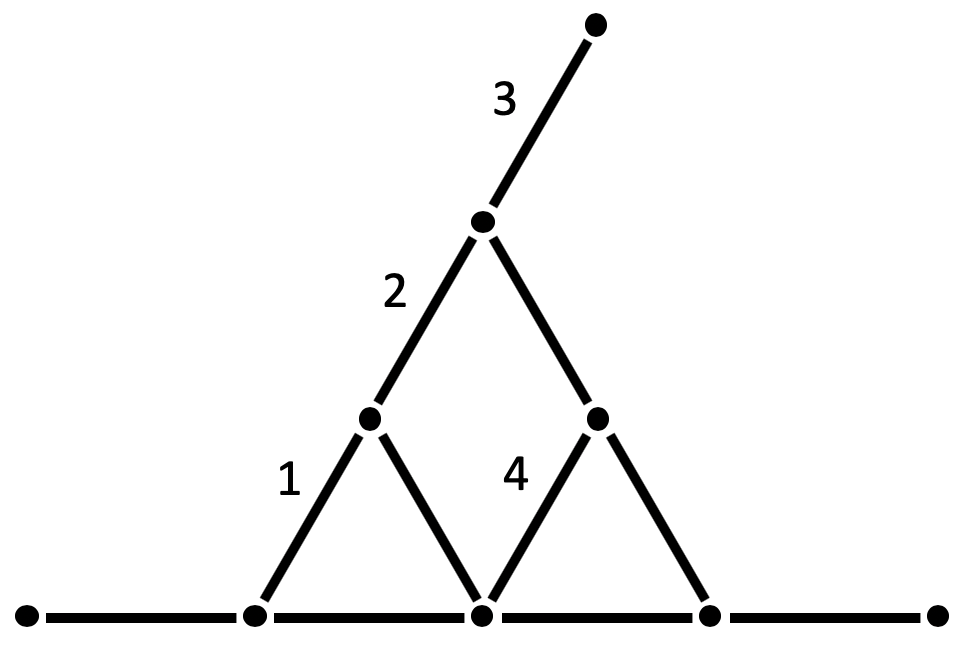}
\caption{Example of a redundant array of antennae.}
\label{fig:redundant_array}
\end{figure}

First, only a limited number of triangles can be constructed from the array of antennae of a telescope, therefore some information will be lost compared to simple baseline measurements. \citet{monnier_phases_interferometry} count that for an array made of $N$ antennae, there are $n = \binom{N}{3} = N(N-1)(N-2)/6$ possible closed triangles, $\binom{N}{2}$ independent Fourier phases and $\binom{N-1}{2}$ independent closure phases. Therefore the amount of phase information recovered from closure phases is
\begin{equation}
    \frac{\binom{N-1}{2}}{\binom{N}{2}}=\frac{(N-1)(N-2)}{2}\times \frac{2}{N(N-1)} = \frac{N-2}{N} = 1- \frac{2}{N}.
\end{equation}
With as little as 40 antennae, we are able to recover $95\%$ of the phase information but 1000 antennae are not enough to reach $99.9\%$. Note that for the 296 antennae of the SKA1-\textit{Low} central array and the 48 antennae of LOFAR, we recover respectively $99.3\%$ and $95.8\%$ of the phase information from closure phases. However, even if almost all the phase information is recovered, \citet{readhead_1988} find that the noise level of a phase-only observation will still be at least twice higher than a map made from full visibility data, because when ignoring amplitudes, half of the signal is lost. Additionally, \citet{readhead_1988} show that because the bispectrum is a triple product, there will be new sources of noise compared to single baseline observations. First, if the same signal is measured on different time intervals, the observed bispectrum will not only be the product of three complex numbers $V_{ij}$ anymore, but of the sum of each observation on each time interval: 
\begin{equation}
\mathcal{B}_{ijk} = \left( \sum_{m} V_{ij,\tau_m} \right) \times \left( \sum_{m} V_{jk,\tau_m} \right) \times \left( \sum_{m} V_{ki,\tau_m} \right),
\end{equation}
where $V_{ij,\tau_m}$ is the visibility measured for baseline $ij$ on time interval $\tau_m$. Then the cross terms combining signals integrated on different time intervals (e.g. $V_{ij,\tau_m}$ and $V_{jk,\tau_l}$) will give incoherent phase terms that can be assimilated to noise. The second potential source of noise mentioned by \citet{readhead_1988} corresponds to the same kind of reasoning, but in the spatial domain: if we consider the triangle formed by three baselines $(ij,ik,kl)$, then on a redundant array there will be contributions not only from the $ij + jk + kl = 0$ triangles but also from identical baselines who are not part of a triangle. See the example on Fig. \ref{fig:redundant_array} of a redundant array: all numbered vectors correspond to the same baseline but only 1, 2 and 4 form triangles. When probing $ijk$ triangles, the measured bispectrum will take the sum of the four signals as the visibility corresponding to this baseline (e.g. $V_{ij}=V_1+V_2+V_3+V_4$) and additional cross terms, irrelevant to closure phases because they include $V_3$, will arise.
\citet{readhead_1988} show however than these two types of noise can be reduced to a sensible signal-to-noise ratio if enough frames are used in the integration.
Similarly, \citet{carilli_2018_bispectrum} mention that some instrumental effects such as polarisation leakage or cross coupling of antennae, called "closure errors" can lead to a departure from the closure relation. 

Finally, the limited number of triangles one can construct from $N$ given antennae will limit the sampling of $(u,v)$ space and worsen the sparsity of observations. Applying our method to sparse data goes beyond the scope of this work, but it would be interesting to see how triangle correlations perform with this additional difficulty. For now, we will limit ourselves to noisy data sets. We refer the interested reader to \citet{trott_watkinson_2019}, where the authors compute the normalised bispectrum defined in \citet{bispectrum_xray_2019} from the closed triangle visibilities of MWA Phase II.

\subsection{Instrumental effects}
\label{subsec:noise}

\begin{figure*}
\centering
\includegraphics[width=0.95\textwidth]{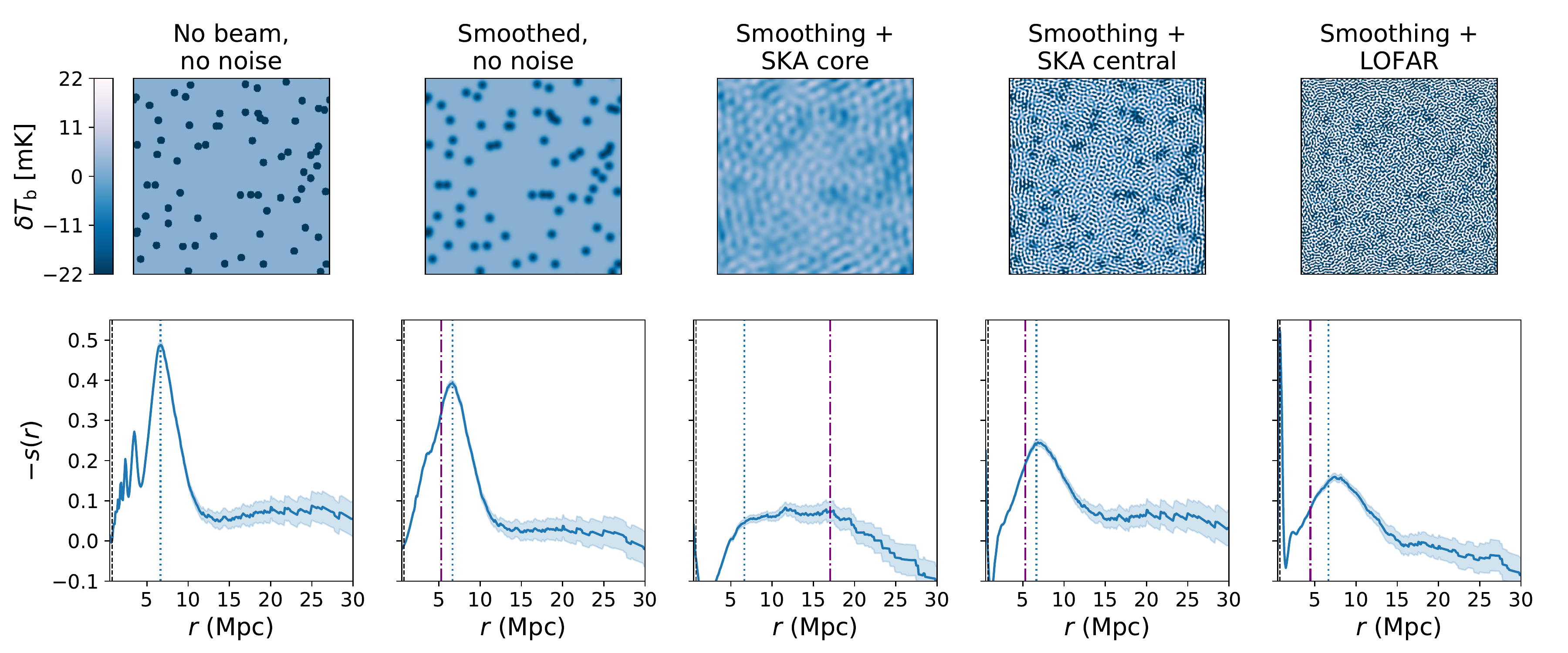}
\includegraphics[width=0.95\textwidth]{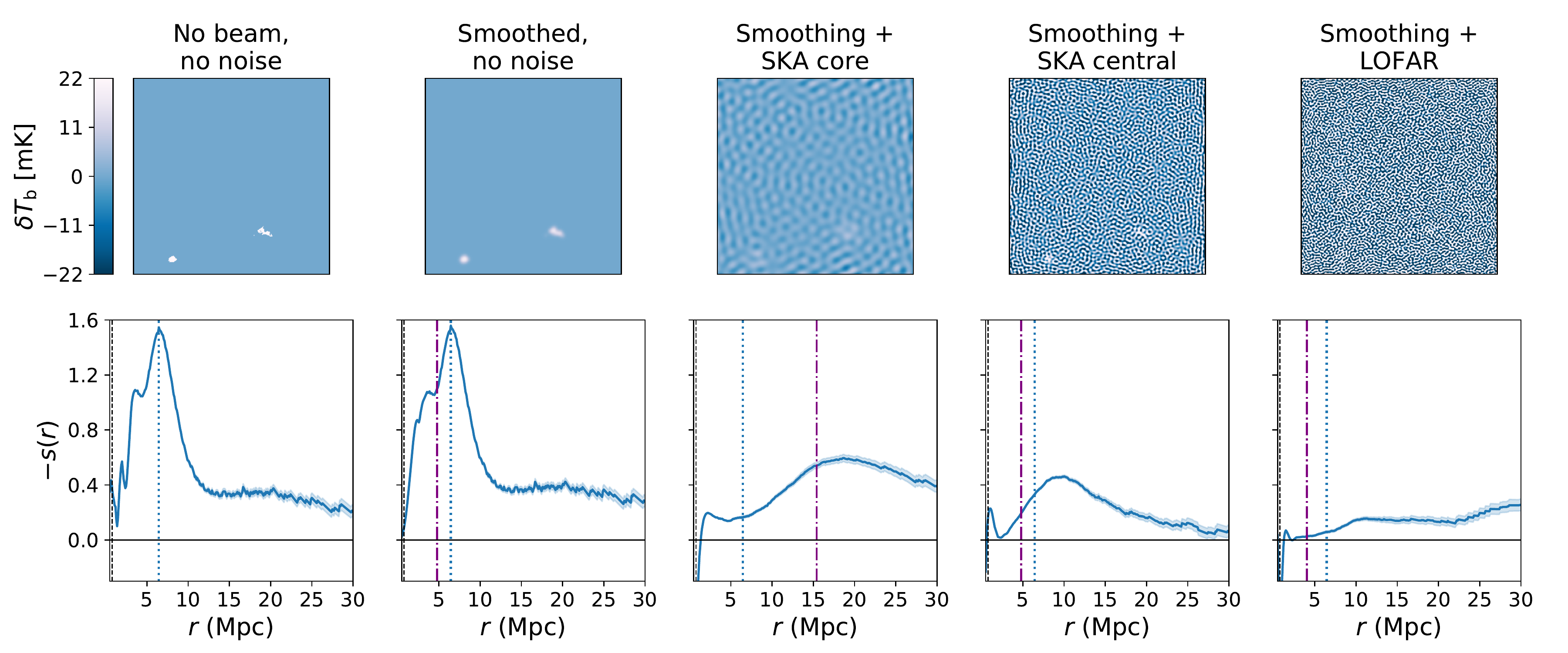}
\includegraphics[width=0.95\textwidth]{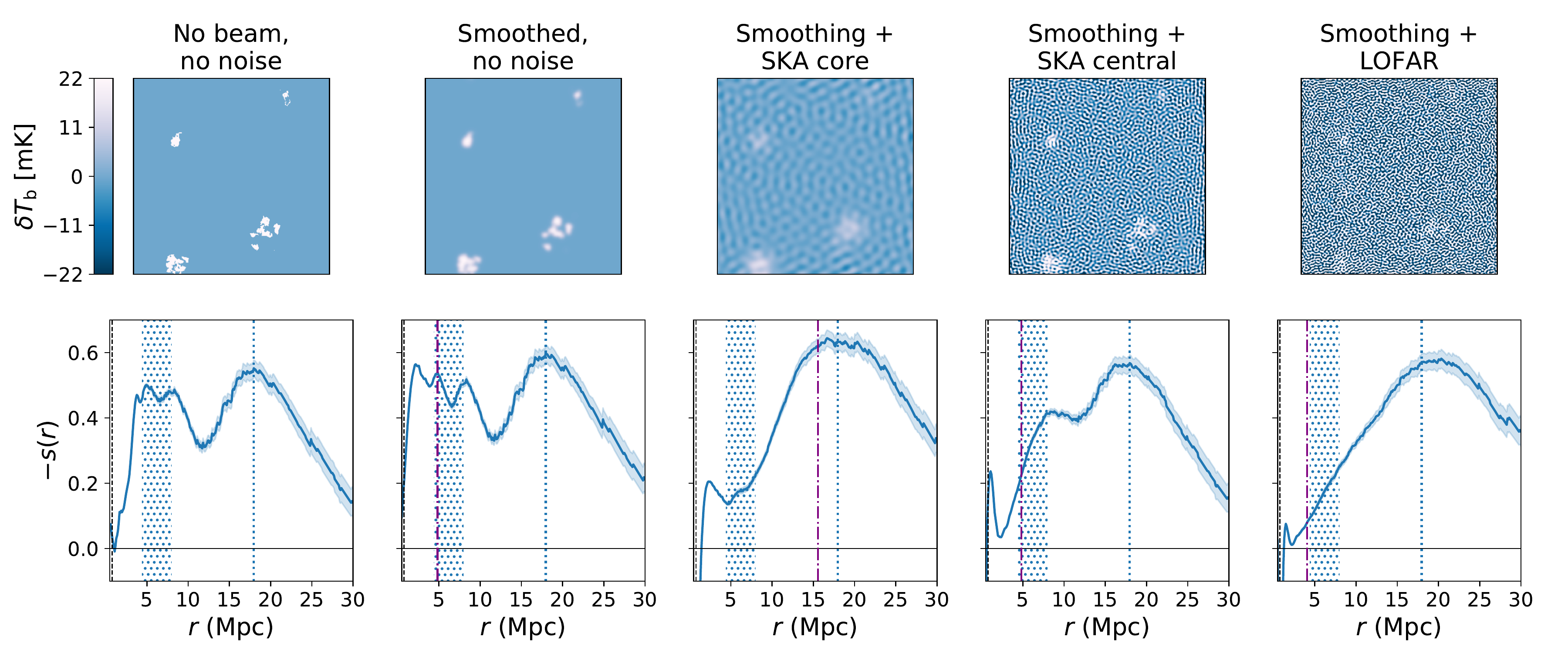}
\caption{Comparison of phase correlations for a 2D brightness temperature map with dimensions $N=512$, $L=400~\mathrm{Mpc}$ for different types of instrumental noise. From left to right: the clean signal, the field smoothed by a Gaussian beam corresponding to the angular resolution of SKA central, and the clean field with added smoothing and noise from each experiment, respectively SKA1-Low core, SKA1-Low central and LOFAR. In each case, the integration time is 1000 hours. Note that $s(r)$ was computed on $\delta T_\mathrm{b}$ maps, derived from the neutral field, hence the signal is mostly negative; for readability here we have represented $-s(r)$. The dotted blue lines (regions) indicate the peaking scale found on the clean field (left panel) and the purple dash-dotted lines the smoothing scale of the corresponding experiment. From top to bottom: toy model of 70 $R=10\mathrm{px}$ binary bbbles, 21CMFAST simluation at $z=6.2$ and $z=6.4$ respectively. Error bars correspond to the variance estimated from a Gaussian random field of same dimensions.}
\label{fig:noise_comparison}
\end{figure*}

To see how our method performs when applied to actual observations, we now add instrumental effects to our ionisation maps: beam smoothing and noise corresponding to observations by the core of SKA1-\textit{Low}, its central area, and by LOFAR.
We pick three simulated ionisation fields: the first comes from the toy models described in Section \ref{sec:bubble_boxes}. It is made of 70 bubbles of radius $R=10~\mathrm{px}$, and assumed to correspond to a redshift $z=9$. The two others are extracted from 21CMFAST at redshifts $z=6.2$ and 6.4. They are shown, in this order, on the left panels of Fig. \ref{fig:noise_comparison}. We first convert each one of them into a brightness temperature map according to Eq. \ref{eq:brithness21cm} with the following cosmology: $\Omega_m=0.309$, $\Omega_b=0.049$ and $H_0 = 67.74~\mathrm{km.s}^{-1} \mathrm{Mpc}^{-1}$. 

The resolution of 21-cm tomographic data will be first limited by the angular resolution of the interferometer considered. The full width at half maximum (FWHM) of an interferometer is given by (in radians):
\begin{equation}
\theta_\mathrm{AR} = \frac{\lambda}{b_\mathrm{max}},
\end{equation}
where $\lambda$ is the redshifted 21cm wavelength i.e. $\lambda = 21\mathrm{cm} \times (1+z)$ and $b_\mathrm{max}$ is the maximum baseline of the interferometer. We have $b_\mathrm{max}= 3500$, $1000$ and $3400~\mathrm{m}$ for LOFAR, SKA1-\textit{Low} core and SKA1-\textit{Low} central respectively. 
To account for this effect, we convolve the $\delta T_b$ map with a Gaussian kernel of FWHM $\theta_\mathrm{AR} d_\mathrm{c} (z)$, with $d_\mathrm{c}(z)$ the comoving distance at the redshift (i.e. frequency) considered. For SKA central, we get the second-to-left panels on the figure -- note that because they have similar maximum baselines, the angular resolution of SKA central is close the one of LOFAR. For SKA core, because $b_\mathrm{max}$ is much smaller, the smoothing blurs the shape of the ionised bubbles to an extreme point, and our method will perform poorly. Finally, we simulate realistic instrumental noise by using the measurement equation software OSKAR\footnote{\url{https://github.com/OxfordSKA/OSKAR}}.

\begin{table}
	\centering
	\caption{Peaking scales (Mpc) for a toy model with different types of observational effects: smoothing due to angular resolution and noise for an integration time of $1000~\mathrm{h}$. Actual radius size is $7.8~\mathrm{Mpc}$.}
	\label{tab:noise}
	\begin{tabular}{lcccc} % four columns, alignment for each
		\hline
		 & Clean & Smoothed & Smoothing &  Smoothing\\
          & signal & signal & + 1000h & scale \\
		\hline
		LOFAR & 6.7 & 6.6 & 7.4 & 4.5\\
		SKA1-\textit{Low} core & 6.7 & 8.2 & -- & 17.0\\
		SKA1-\textit{Low} central & 6.7 & 6.7 & 6.9 & 5.3\\
		\hline
	\end{tabular}
\end{table}

Fig. \ref{fig:noise_comparison} presents the results: the first column corresponds to triangle correlations for a clean field. We identify the peaking scale for this clean field in order to compare it to the scale picked on corrupted data later on and indicate it as a dotted blue line on each plot. Note that for the third field (21CMFAST at $z=6.4$), there are two scales picked up. The larger one is well defined, whereas we use an interval $6.2 \pm 1.6~\mathrm{Mpc}$ for the smaller one. On the figure, the second column is for the field smoothed with the angular resolution of SKA1-\textit{Low} central; and the last three for the smoothed signal with instrumental noise from each of the three experiments considered (respectively SKA1-\textit{Low} core, central and LOFAR) and 1000 hours of integration time. We choose 1000h as it is the most commonly used value in the literature, but recent works have shown that as few as 324h of observations with SKA can be sufficient to differentiate between different reionisation models \citep{binnie_2019}. The dash-dotted lines on each plot mark the smoothing scale of the corresponding experiment, whose values are given in Table \ref{tab:noise}. One can see that results with SKA core are not satisfying because of a too low angular resolution that blurs the edges of the ionised regions: the signal is mostly flat or seems to pick up the smoothing scale. However, results for SKA central and LOFAR prove very satisfactory on the toy model (first row), with a clear peak at a scale close to the one found with clean signal (see Table \ref{tab:noise} for details): applying the linear relation of Eq. \ref{eq:rvsR_binary} to the peaking scale could give a good approximation of the actual size of ionised regions in the image. When the integration time is increased, we get for both LOFAR and SKA1-\textit{Low} central a peaking scale even closer to the one found for clean signal. Unfortunately, the LOFAR sensitivity does not allow to extract much information from the 21CMFAST maps: at $z=6.2$ the signal is mostly flat, and although there is a clear peak for the $z=6.4$ map, it cannot resolve the two characteristic scales, however picked up when accounting for SKA central sensitivity. Note that for each plot, the error bars correspond to the variance of the TCF computed for a GRF of same dimensions.
We have also tried our statistic on 21CMFAST boxes at higher redshifts, and we find that, because there is no clear characteristic scale in the field (see Section \ref{sec:21cmFAST}), the TCF peaks at the smoothing scale corresponding to the telescope considered (listed in Table \ref{tab:noise}). There is however no risk of misinterpretation of this peak since real data would be deconvolved from telescope properties such as angular resolution before being analysed.

In all the work above, we assumed that foreground pollution was completely removed from the data cubes: \citet{chapman_2015_foregrounds_aska} presented efficient foreground removal techniques that will produce good quality 21cm maps. However, foreground residuals could still impact our results. A precise instrumental calibration is usually required to separate the 21cm signal from foregrounds and is one of the main challenges of upcoming experiments although as mentioned before, it seems that bispectrum phases and therefore our results would be unaffected by calibration errors \citep{thyagarajan_2018_bispectrum_phases}.

\section{Why phase information?}
\label{sec:why_phases}

\begin{figure*}
\centering
\includegraphics[width=0.95\textwidth]{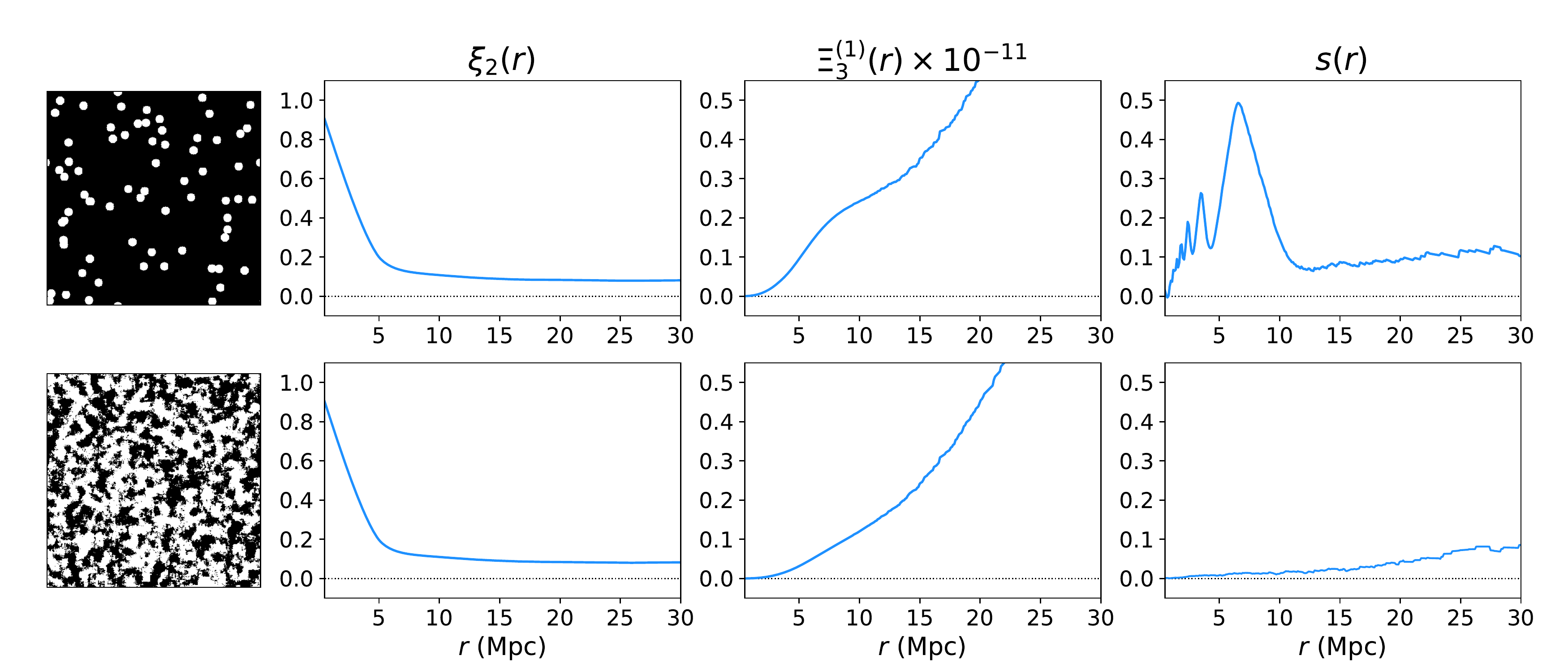}
\caption{Comparison of results on phase correlations for two ionisation fields with identical power spectra and dimensions ($N=512$, $L=400~\mathrm{Mpc}$) but different phase information: lower panels correspond to the field from upper panels after having reshuffled the Fourier phases. Left panels show the 2D ionisation field in real space, middle left, middle right and right panels respectively show the corresponding 2-PCF $\xi_2(r)$, the scaled modified 3-PCF $\Xi_3^{(1)} (r)$ (see Eq. \ref{eq:3PCF_full_bispectrum}), and the triangle correlation function $s(r)$.}
\label{fig:same_2PCF}
\end{figure*}

\begin{figure}
	\centering
	\includegraphics[width=.9\columnwidth]{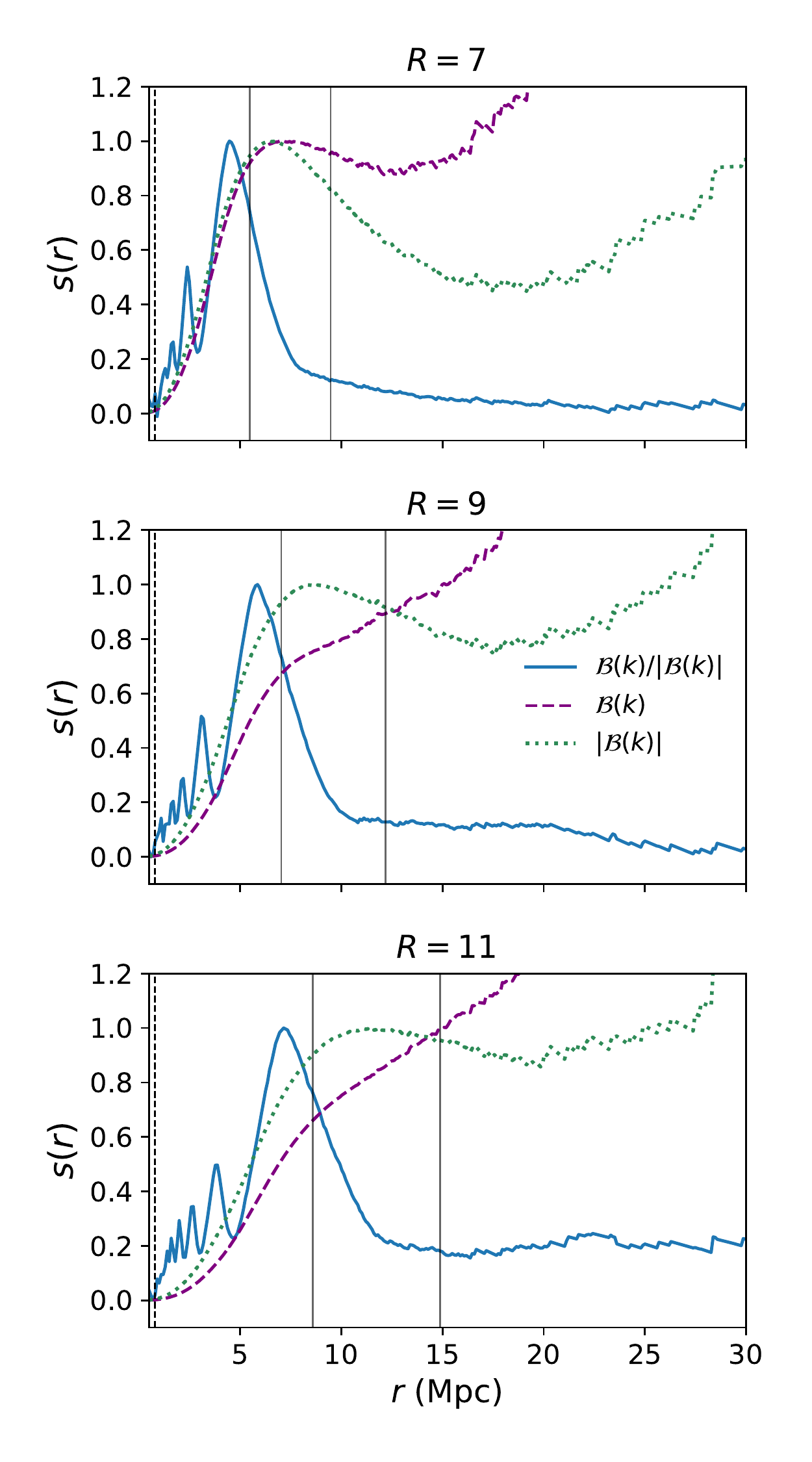}
    \caption{Triangle correlations computed on a box with dimensions $N=512$ and $L=400~\mathrm{Mpc}$, filled with 50 binary bubbles of radius $R$. Upper panel corresponds to $R=7~\mathrm{px}$, middle to $R=9~\mathrm{px}$ and lower to $R=11~\mathrm{px}$. Each time, the vertical lines correspond to $R$ and $\sqrt{3}R$. The blue solid line is the triangle correlation function as defined before. The purple dashed line corresponds to a triangle correlation function computed for the full bispectrum rather than only its phase factor $\Xi_3^{(1)} (r)$ and the green dotted line to triangle correlations computed from the amplitude of the bispectrum only $\Xi_3^{(2)} (r)$ (see text for details).}
\label{fig:SC_with_without_amplitude}
\end{figure}

In order to see if our phase-only estimator truly supplements power spectrum information, we generate a 2D ionisation field made of 70 binary disks of radius $R=10~\mathrm{px}=7.8~\mathrm{Mpc}$ and compute its 2-point correlation function and its TCF. We then shuffle its Fourier phases -- replace them by random phases ranging from 0 to $2\pi$, and compute the corresponding 2-PCF and TCF. Results can be seen on the upper and lower panels of Fig. \ref{fig:same_2PCF} respectively. On the left panels, we see that reshuffling the phases has made the field lose all its structure: there are no bubbles any more. Because we kept the absolute value of the field unchanged, the 2-PCFs in the second column are exactly identical. However, the extra information carried by phases is clearly visible when comparing the TCFs on the right-hand panels: there is almost no signal for the field with random phases whereas the field with structured phases exhibits a clear peak.

For completeness, we compute three different types of triangle correlations encompassing more or less phase information and compare results on Fig. \ref{fig:SC_with_without_amplitude}. We do so for three boxes filled with 70 binary bubbles: the first (upper panel) with bubbles of radius $R=7$, the second (middle panel) $R=9$ and the third (lower panel) $R=12$.
\begin{enumerate}
\item The first type of triangle correlations considered is the one we have used so far, defined in Eq. \ref{eq:def_triangle_correlation_function} as a modified inverse Fourier transform of the phase factor of the bispectrum (blue solid line on Fig. \ref{fig:SC_with_without_amplitude}).
\item The second uses the full bispectrum, amplitude included (purple dashed line on Fig. \ref{fig:SC_with_without_amplitude}):
\begin{equation}
\label{eq:3PCF_full_bispectrum}
\Xi_3^{(1)} (r) = \left( \frac{r}{L} \right) ^{3D/2} \sum_{k,q \, \leq \, \pi/r} \omega_D \left(pr\right) \, \mathcal{B} \left( \bm{k}, \bm{q} \right) .
\end{equation}
\item The third only uses the amplitude of the bispectrum (green dotted line on Fig. \ref{fig:SC_with_without_amplitude}):
\begin{equation}
\label{eq:3PCF_amplitude_only}
\Xi_3^{(2)} (r) = \left( \frac{r}{L} \right) ^{3D/2} \sum_{k,q \, \leq \, \pi/r} \omega_D \left(pr\right) \, \vert \mathcal{B} \left( \bm{k}, \bm{q} \right) \vert.
\end{equation}
\end{enumerate}
We see that the main advantage of the phase-only TCF is that, compared to statistics using bispectrum amplitude, the peaking scale is much better defined. It is also less sensitive to the filling fraction than the two other functions: in the bottom panel, where bubbles are larger and the filling fraction higher, phase correlations still exhibit a well-defined peak whereas the other two flatten. 
% For completeness, we compute the three types of triangle correlations mentioned before for two boxes: one filled with 20 binary disks of radius 14 and 60 of radius 4, and one filled with enough binary ellipses of axes (4,14) to reach the same filling fraction as the former. For $\Xi_3^{(1)}$ and $\Xi_3^{(2)}$, we find the signal to be perfectly identical to the results shown on Fig. \ref{fig:SC_with_without_amplitude} for bubbles. Therefore, the phase-only triangle correlation function is the only estimator able to differentiate between ellipses and bubbles of two different sizes.

\section{Computational performance}
\label{sec:computational_perf}

In Sec. \ref{sec:21cmFAST} we compared the results of different methods to derive a bubble size distribution or equivalent, but we did not mention computational performance. Note that our code has been parallelised via OpenMP \citep{openmp13} in order to reduce computing times: because most of the code consists of sums, i.e. nested loops, the parallelisation is very efficient. Three parameters will play an essential role in determining the computing time necessary to evaluate triangle correlations:
\begin{enumerate}
    \item The number $N$ of pixels in the box.
    \item The range of correlations scales $r$ for which we compute the triangle correlations: because of the limits of our sum ($\sum_{k \leq \pi / r}$, see Eq. \ref{eq:def_triangle_correlation_function}), there will be more terms to sum over for smaller scales. Therefore, in this work, we have chosen to compute triangle correlations for $0.5~\mathrm{Mpc} \leq r \leq 30~\mathrm{Mpc}$.
    \item The box side length $L$, because it determines the sampling of $k$-cells: for smaller $\Delta k$, there will be more modes with norm smaller than $\pi / r$ and hence more terms to sum over.
\end{enumerate}
Most of the boxes analysed in this work had $512^2$ pixels for a side length $L=400~\mathrm{Mpc}$ and triangles correlations were computed for $r/\mathrm{Mpc} \in [0.5 , 30]$, for an average computation time of 180 minutes on 20 cores. For comparison, the SPA algorithm, as implemented by \citet{giri_2018_bubble_sizes}, takes about half an hour to run on a 3D box with same side length and sampling; whereas RMFP, as implemented in 21CMFAST, only takes a couple minutes. Therefore more work is required to improve the computational efficiency of our algorithm further and so its usability. An approach similar to the one presented in the Appendix of \citet{eggemeier_2015}, where the nested sums in Eq. \ref{eq:def_triangle_correlation_function} are replaced by rotational averages in real space, could greatly improve computation times.

\section{Conclusions}
\label{sec:conclusion}

Following the work of \citetalias{obreschkow_2013}, we have constructed a new statistical tool, based on phase information only and called the triangle correlation function (TCF), which can be used to determine the characteristic scale of ionised regions on 21cm interferometric data from the Epoch of Reionisation. Indeed, if we plot this function over a range of correlation scales for simple fields, made of perfectly spherical fully ionised regions on a neutral background, we find a peaked signal, and the peaking scale can be directly related to the actual size of the bubbles. From such toy models, we have derived important properties for the TCF: it can differentiate between ionised regions of different sizes, and distinguish spherical from elongated structures. We have also found its results to be more reliable on scales smaller than about a twentieth of the physical side length of the field studied, as the finite size of the box implies more sample variance at larger scales. To see if our method can be applied to observational data, we have confronted the TCF with ionisation fields corrupted by instrumental noise or angular resolution from LOFAR or SKA, and we have found that it still performs well at giving the characteristic scale of ionised regions, as long as the integration time is sufficient -- here, we have worked with 1000 hours. By comparing results for 3-PCF including amplitude information to $s(r)$, we also proved that a statistical tool using phase information only will be more efficient at picking up the characteristic scale of spherical structures in a field than a correlation function also using information from the bispectrum amplitude. Indeed, the phase-only function is more peaked, allowing a better identification of the characteristic scale. 
Moving on to the more elaborate reionisation simulation 21CMFAST \citep{21cmFAST_2007}, we have found that our method gives a good estimation of the size of remote neutral islands at the very end of the reionisation process. In particular, and contrarily to other BSD algorithms such as RMFP and SPA, it is able to resolve two characteristic scales in a field. In the early stages of the simulation, because there are many very small ionised regions covering the neutral background, the signal is dominated by Poisson noise, and there is no clear characteristic scale to pick up. Therefore, for fields with a very low global ionised fraction, we rather use our method to infer an upper limit on the size of ionised regions. Note that during the \textit{overlap} phase, positive correlations from ionised regions overlap with negative signal from neutral zones and the signal flattens out: an absence of signal can be interpreted as reionisation being in its middle stages and so we can learn about the duration of the process. In general, it will be more difficult to extract phase information from non binary ionisation field because of the more complex structure  of the field, which is reverberated in the phase information: Fig. \ref{fig:epsilon_vs_realbox} compares the real-space phase information of one of our toy models with $\epsilon (\bm{r})$ for one of our 21CMFAST boxes. These results, on both toy models and more elaboration simulations, hold when instrumental effects such as telescope angular resolution and instrumental noise are added to clean maps.

Apart from noise and instrumental resolution, there will be some other important questions to solve before applying our method to true 21cm tomographic data. 
\citet{kakiichi_2017_HII_tomography} proved that the cold spots in 21cm tomographic images trace $\ion{H}{II}$ regions more efficiently at low redshift ($z \leq 7$) and high filling fraction ($x_\ion{H}{II} \geq 0.4$): earlier in the reionisation process, cold spots due to a local underdensity rather than ionisation are more frequent (see Sec. \ref{sec:21cm}). Corroborated with the fact that the TCF performs well on remote neutral islands at the end of the reionisation process, we can expect to get good results on low redshift data. However, correctly identifying ionised regions in 21cm observations i.e. transforming a map of the differential brightness temperature $\delta T_b$ into a binary field made of fully ionised and fully neutral regions remains a challenge -- for an extended discussion of this issue, see \citet{giri_2018_bubble_sizes}. Note that this is not an issue for the friends-of-friends \citep{iliev_2006,friedrich_2011_topology_tomography} and watershed methods \citep{lin_2016_review} since they segment the data themselves. 
Additionally, it remains to be seen exactly how to exploit closure phases, and what is the effect of the quality of the data, especially its sparsity, on our results. We would also need to see how foreground removal techniques impact the non-Gaussianity of the signal, and therefore our results. Finally, the practical application of triangle correlations to observational data would require forward-modelling i.e. comparing measurements to the signal obtained for a number of simulations corresponding to various parametrisations of reionisation to infer the reionisation scenario corresponding to the observation. In this perspective, one would need to compute the triangle correlations of many more simulations.

\section*{Acknowledgements}
The authors thank the referee for a useful report which helped improve the results of this paper. They also thank Catherine A. Watkinson, Marian Douspis and Beno\^{i}t Semelin for useful discussions on various aspects of the analysis; Nithyanandan Thyagarajan for his helpful comments on a draft version of this paper; and Emma Chapman for kindly providing results from OSKAR runs to perform our noise analysis. AG and JRP acknowledge financial support from the European Research Council under ERC grant number 638743-FIRSTDAWN. AG's work is supported by a PhD studentship from the UK Science and Technology Facilities Council (STFC).

This research made use of \texttt{astropy}, a community-developed core Python package for astronomy \citep{astropy,astropy2}; \texttt{matplotlib}, a Python library for publication quality graphics \citep{hunter_2007}; \texttt{scipy}, a Python-based ecosystem of open-source software for mathematics, science, and engineering \citep{scipy} -- including \texttt{numpy} \citep{numpy}, and \texttt{emcee}, an implementation of the affine invariant MCMC ensemble sampler \citep{emcee}.

%%%%%%%%%%%%%%%%%%%%%%%%%%%%%%%%%%%%%%%%%%%%%%%%%%


\bibliographystyle{mnras}
\bibliography{bibliography}

%%%%%%%%%%%%%%%%%%%%%%%%%%%%%%%%%%%%%%%%%%%%%%%%%%

\appendix

\section{Analytical derivation for a toy model}
\label{appendix:derivation}

Consider a 2D box of volume $V=L^2$ filled with $n$ fully ionised bubbles of radius $R$, randomly distributed throughout the box so that their centres are located at $ \bm{a} _i$ for $i \in \{ 1, n \}$. 
The ionisation field of the box is then
\begin{equation}
x_\ion{H}{II}(\bm{r}) = \sum_{i=1}^n \Theta \left( \frac{ \vert \bm{r} - \bm{a}_i \vert}{R} \right),
\end{equation}
where $\Theta \left( x\right)$ is the Heaviside step function worth 1 if $x \leq 1$ and 0 otherwise. 
We need to take the Fourier transform of this field, derived as follows:
\begin{equation}
\begin{aligned}
\hat{x} ( \bm{k} ) & = \frac{1}{V} \iint \sum_{i=1}^{n} \Theta \left( \frac{ \vert \bm{r} - \bm{a}_i \vert}{R} \right) \mathrm{e}^{- i \bm{k} \cdot \bm{r}} \mathrm{d}^2 \bm{r} \\
& = \frac{1}{V} \sum_{i=1}^{n} \mathrm{e}^{- i \bm{k} \cdot \bm{a}_i} \int_{r=0}^{R}  r \mathrm{d}r  \int_{\theta=0}^{2\pi} \mathrm{d}\theta \, \mathrm{e}^{- i kr \mathrm{cos}\theta}  \\
& = \frac{1}{V} \sum_{i=1}^{n} \mathrm{e}^{- i \bm{k} \cdot \bm{a}_i} \int_{r=0}^{R}  r \mathrm{d}r  \int_{-\pi}^{\pi} \mathrm{d}\theta \, \mathrm{e}^{i kr \mathrm{cos}\theta}  \\
& = \frac{1}{V} \sum_{i=1}^{n} \mathrm{e}^{- i \bm{k} \cdot \bm{a}_i} \int_{r=0}^{R}  r \mathrm{d}r \times 2 \pi J_0(kr) \\
& = \frac{2\pi}{k^2 V} \sum_{i=1}^{n} \mathrm{e}^{- i \bm{k} \cdot \bm{a}_i} \int_{0}^{kR}  z J_0(z) \, \mathrm{d}z \\
& = \frac{2\pi R}{k V} J_1(kR) \sum_{i=1}^{n} \mathrm{e}^{- i \bm{k} \cdot \bm{a}_i} .
\end{aligned}
\end{equation}
where to go from the first to the second line, we have used the change of variables $\bm{r} = \bm{r} - \bm{a}_i$. From the fourth to the fifth, we have let $z=kr$ to be able to use:
\begin{equation}
\int_0^v w J_0 (w) \, \mathrm{d}w = v J_1(v).
\end{equation}
If we define $W(y)=J_1(y)/y$, the Fourier transform of the top-hat window function in 2D, we have the final expression of the Fourier transform of our ionisation field:
\begin{equation}
\hat{x} ( \bm{k} ) = \frac{2\pi R^2}{V} W(kR) \sum_{i=1}^{n} \mathrm{e}^{- i \bm{k} \cdot \bm{a}_i} \simeq \frac{2\bar{x}_\ion{H}{II}}{n}  W(kR) \, \sum_i^n \mathrm{e}^{- i \bm{k} \cdot \bm{a}_i}.
\end{equation}
Indeed, if we let $\rho$ the mean number density of ionised bubbles such that $\rho = n / V$ and ignore overlapping\footnote{This will be a reasonable assumption for the filling fractions considered.}, $\bar{x}_\ion{H}{II} = \pi R^2 \rho$.
From this we deduce analytic expressions for the power spectrum
\begin{equation}
\label{eq:PS_toy_model}
\begin{aligned}
\mathcal{P}(\bm{k})  & = \vert \hat{x} ( \bm{k} ) \vert ^2 \\ 
&  = 4 \pi^2 \, W(kR)^2 \, \left( \frac{R}{L} \right)^4  \,  \sum_{i,j=1}^n \mathrm{e}^{- i \bm{k} \cdot \left( \bm{a}_i - \bm{a}_j \right) } \\ 
& = 4 \pi^2 \, W(kR)^2 \, \left( \frac{R}{L} \right)^4  \times 2 \sum_{i \leq j} \mathrm{cos} \left[ \bm{k} \cdot \left( \bm{a}_i - \bm{a}_j \right) \right] ,
\end{aligned}
\end{equation}
which is a real number; and the bispectrum of the box
\begin{equation}
\label{eq:bispectrum_toy_model}
\begin{aligned}
 \mathcal{B}(\bm{k},\bm{q}) & = \hat{x} ( \bm{k} ) \, \hat{x} ( \bm{q} ) \, \hat{x} ( \bm{-k-q} ) \ldots \\
 & = 8 \pi^3 \ W(kR) \, W(qR) \, W\left(  \lvert \bm{k+q} \rvert R \right) \, \left( \frac{R}{L} \right)^6    \\ & \times \ \sum_{\alpha, \beta, \gamma = 1}^n \mathrm{e}^{- i \left[ \bm{k} \cdot \left( \bm{a}_\alpha - \bm{a}_\gamma \right) + \bm{q} \cdot \left( \bm{a}_\beta - \bm{a}_\gamma \right) \right] } .
\end{aligned}
\end{equation}
To find the analytic expression of our triangle correlations function for this toy model, we then plug Eq. \ref{eq:bispectrum_toy_model} into Eq. \ref{eq:def_triangle_correlation_function}.

\begin{figure}
	\centering
	\includegraphics[width=0.75\columnwidth]{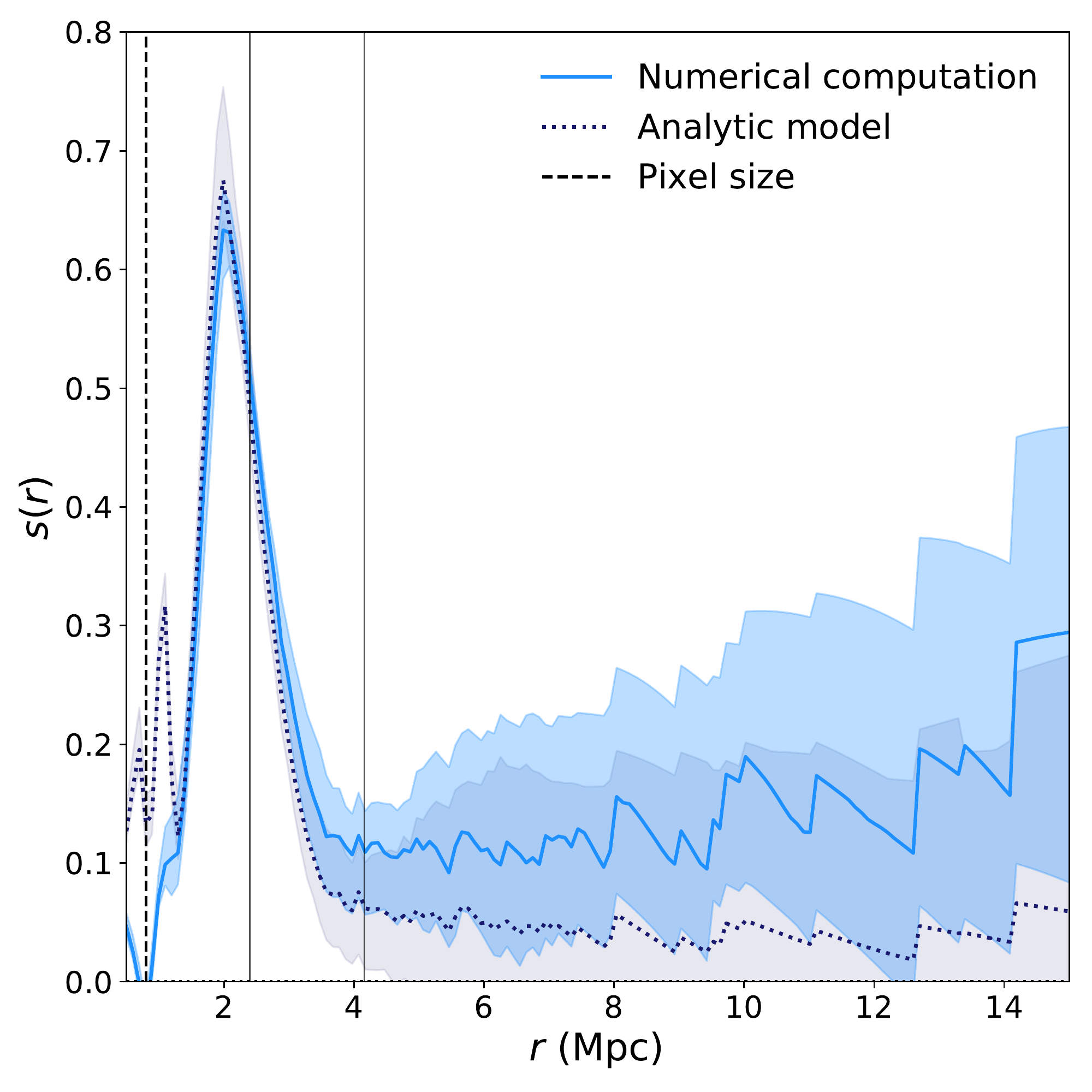}
    \caption{Triangle correlation function computed from the numerical expression of the bispectrum in Eq. \ref{eq:def_bspectrum} (solid line) and from the analytic expression of the bispectrum for a known bubble distribution as in Eq. \ref{eq:bispectrum_toy_model} (dotted line) for 20 realisations a box of $100^2$ pixels and side length $L=80~\mathrm{Mpc}$ filled with~20 binary bubbles of radius $R=3~\mathrm{px}=2.4~\mathrm{Mpc}$. Vertical lines correspond to $R$ and $\sqrt{3}R$.}
    \label{fig:toy_model}
\end{figure}

In Fig. \ref{fig:toy_model}, we compare the result of a numerical integration of the triangle correlation function computed from the bispectrum in Eq. \ref{eq:def_bspectrum} with a version where we compute analytically the triangle correlation function, knowing the locations of the ionised bubbles and the size of the box only, according to Eq. \ref{eq:bispectrum_toy_model} and detailed above. We consider 20 realisations of a box of dimensions $L=80~\mathrm{Mpc}$ for $100^2$ pixels, filled with 20 bubbles of radius $R=3~\mathrm{px}=2.4~\mathrm{Mpc}$\footnote{The choices of box size and bubble number are limited by the computational cost of the analytic method.}. We see that there is a good match between both computational methods: we have the confirmation that our estimator indeed traces the bubble distribution in the ionisation field considered.

% Don't change these lines
\bsp	% typesetting comment
\label{lastpage}
\end{document}